\documentclass[11pt]{article}
\usepackage{amssymb}
\usepackage{amsfonts}
\usepackage{amsmath}
\usepackage{mathrsfs}
\usepackage{eufrak}
\usepackage{fancyhdr}
\usepackage[dvips]{graphicx}

\catcode`@=11
\renewcommand{\section}{\@startsection{section}{1}{0pt}{\medskipamount}
{\medskipamount}{\large\bf}}
\numberwithin{equation}{section}
\catcode`@=12

\def\a{\alpha}

\def\g{\gamma}

\def\eps{\epsilon}

\def\z{\zeta}
\def\h{\eta}
\def\t{\theta}

\def\la{\lambda}
\def\m{\mu}
\def\n{\nu}
\def\r{\rho}
\def\s{\sigma}
\def\p{\phi}
\def\vp{\varphi}
\def\c{\chi}

\def\Ga{\Gamma}
\def\Ups{\Upsilon}

\def\sfrac#1#2{{\textstyle\frac{#1}{#2}}}
\def\rd#1{\buildrel{_{_{\hskip 0.01in}\rightarrow}}\over{#1}}
\def\ld#1{\buildrel{_{_{\hskip 0.01in}\leftarrow}}\over{#1}}
\def\+{\dagger}
\def\={\ =\ }
\def\pa{\partial}
\def\tr{\mathrm{tr}}
\def\det{\mathrm{det}}
\def\res{\mathrm{res}}
\def\ch{\mathrm{ch}}

\def\th{\mathrm{th}}

\newcommand{\beq}{\begin{equation}}
\newcommand{\eeq}{\end{equation}}
\newcommand{\bea}{\begin{eqnarray}}
\newcommand{\eea}{\end{eqnarray}}
\newcommand{\ena}{\end{eqnarray}}
\newcommand{\non}{\nonumber}

\newcommand{\mbf}[1]{{\boldsymbol{#1}}}
\newcommand{\e}{\,\mathrm{e}\,}
\newcommand{\im}{\,\mathrm{i}\,}
\newcommand{\diff}{\mathrm{d}}
\newcommand{\intd}{\mathrm{d}}
\newcommand{\R}{{\mathbb{R}}}
\newcommand{\C}{{\mathbb{C}}}

\newcommand{\Ecal}{{\cal E}}
\newcommand{\mb}{\bar{\mu}}
\newcommand{\Lb}{\bar{L}^{A,B}}
\newcommand{\Lbt}{\widetilde{\bar{L}}^{A,B}}
\newcommand{\Pht}{\widetilde{\Phi}}
\newcommand{\Pt}{\widetilde{P}}
\newcommand{\St}{\widetilde{S}}
\newcommand{\Tt}{\widetilde{T}}
\newcommand{\gt}{\widetilde{g}}
\newcommand{\Psh}{\widehat{\Psi}}
\newcommand{\Sh}{\widehat{S}}

\begin{document}
\begin{titlepage}
\setcounter{page}{0}
\begin{flushright}
ITP--UH--18/04\\
Bicocca-FT-04-05
\end{flushright}

\vspace{5mm}

\begin{center}

{\Large\bf Integrable Noncommutative Sine-Gordon Model}

\vspace{10mm}

{\large Olaf Lechtenfeld} \\
{\em Institut f\"ur Theoretische Physik, Universit\"at Hannover \\
     Appelstra\ss{}e 2, D-30167 Hannover, Germany}\\
{\tt lechtenf@itp.uni-hannover.de}
\\[5mm]
{\large Liuba Mazzanti} \\
{\em Dipartimento di Fisica, Universit\`a degli studi di Milano-Bicocca and\\
     INFN, Sezione di Milano, piazza della Scienza 3, I-20126 Milano, 
 Italy}\\
{\tt liuba.mazzanti@mib.infn.it}
\\[5mm]
{\large Silvia Penati} \\
{\em Dipartimento di Fisica, Universit\`a degli studi di Milano-Bicocca and\\
     INFN, Sezione di Milano, piazza della Scienza 3, I-20126 Milano, 
Italy}\\
{\tt silvia.penati@mib.infn.it}
\\[5mm]
{\large Alexander D. Popov~$^*$} \\
{\em Institut f\"ur Theoretische Physik, Universit\"at Hannover \\
     Appelstra\ss{}e 2, D-30167 Hannover, Germany}\\
{\tt popov@itp.uni-hannover.de}
\\[5mm]
{\large Laura Tamassia} \\
{\em Dipartimento di Fisica Nucleare e Teorica, Universit\`a degli studi 
     di Pavia and \\
     INFN, Sezione di Pavia, via Ugo Bassi 6, I-27100 Pavia, Italy}\\
{\tt laura.tamassia@pv.infn.it}     

\vspace{10mm}

\begin{abstract}
\noindent
Requiring an infinite number of conserved local charges or the existence of
an underlying linear system does not uniquely determine the Moyal deformation 
of 1+1 dimensional integrable field theories. As an example, the sine-Gordon 
model may be obtained by dimensional and algebraic reduction from 2+2
dimensional self-dual U(2) Yang-Mills through a 2+1 dimensional integrable
U(2) sigma model, with some freedom in the noncommutative extension of this
algebraic reduction. Relaxing the latter from U(2) $\to$ U(1) to U(2) $\to$
U(1)$\times$U(1), we arrive at novel noncommutative sine-Gordon equations for 
a {\em pair\/} of scalar fields. The dressing method is employed to construct 
its multi-soliton solutions. Finally, we evaluate various tree-level amplitudes
to demonstrate that our model possesses a factorizable and causal S-matrix 
in spite of its time-space noncommutativity.

\end{abstract}

\end{center}

\vfill

\textwidth 6.5truein
\hrule width 5.cm
\vskip.1in

{\small
\noindent ${}^*$
On leave from Bogoliubov Laboratory of Theoretical Physics, JINR,
Dubna, Russia}

\end{titlepage}

\section{Introduction}

In the low-energy limit string theory with D-branes 
gives rise to noncommutative field theory on the branes
when the string propagates in a nontrivial background with an NS-NS
two-form $B$ turned on \cite{SW}. 
In particular, if the open string has an $N{=}2$ 
worldsheet supersymmetry the tree-level target-space dynamics is described 
by a noncommutative 2+2 dimensional self-dual Yang-Mills (SDYM) theory
\cite{LPS}.
When the target space is filled by $n$ coincident D3-branes the low-energy
model is noncommutative U($n$) SDYM. This theory is integrable 
classically~\cite{T} and in the sense that it possesses a factorized 
S-matrix~\cite{LPS}. 

In the ordinary commutative case it is well known\footnote{
See e.g.~\cite{dimred} and references therein.}    
that dimensional reduction of four dimensional SDYM gives rise 
to many integrable systems in three and two dimensions. This property has been 
shown to extend also to the noncommutative case\footnote{
See \cite{reviews} for reviews on noncommutative field theory.}
where integrable models in three and in two dimensions have been 
constructed \cite{Legare,LPS2,LP1,bieling,wolf,goteborg}. 
In fact, almost all integrable noncommutative models in less than
four dimensions\footnote{
See e.g.~\cite{list1,DMH,list2,murugan,hamanaka,GMPT} and references therein.}
can be obtained in this fashion.

In 1+1 dimensions, noncommutative models have not yet been considered 
extensively, primarily because in this situation time is necessarily
a noncommutative coordinate, which appears to compromise the causality and 
unitarity of the theory \cite{seiberg,gm,AG,CM,GMPT}.
However, one may hope that a huge underlying symmetry improves the situation 
in systems which allow for a Lax-pair formulation and so feature an infinite 
chain of conserved charges.

In this paper, we are searching a noncommutative generalization of the 
sine-Gordon system which, as a hallmark of integrability, possesses a 
well-defined {\em causal\/} and {\em factorized\/} S-matrix.
Furthermore, its equations of motion should admit noncommutative
multi-soliton solutions which represent deformations of the well known 
sine-Gordon solitons.

Previous suggestions for noncommutative sine-Gordon models can be found in 
\cite{GP,CM}. In particular, in \cite{GP} a model was proposed which
describes the dynamics of a complex scalar field by a couple 
of equations of motion. These equations were obtained as flatness conditions
for a U(2) bidifferential calculus~\cite{DMH} and automatically guarantee 
the existence of an infinite number of local conserved currents.
The same equations were also generated in~\cite{GMPT} via a particular 
dimensional reduction of the noncommutative U(2) SDYM equations in 2+2 
dimensions. However, this reduction did {\em not\/} work at the level of 
the action, which turned out to be the sum of two WZW models augmented
by a cosine potential. 
Evaluating tree-level scattering amplitudes it was discovered, furthermore,
that this model suffers from acausal behavior and a non-factorized S-matrix,
meaning that particle production occurs.

At this point it is important to note that the noncommutative deformation
of an integrable equation is a priori not unique, because one may always add
terms which vanish in the commutative limit. For the case at hand, for example,
different inequivalent ans\"atze for the U(2) matrices entering the bicomplex 
construction \cite{DMH} are possible as long as they all reproduce the ordinary
sine-Gordon equation in the commutative limit. It is therefore conceivable
that among these possibilities there exists an ansatz (different from the one 
in \cite{GP,GMPT}) which guarantees the classical integrability of the 
corresponding noncommutative model. What is already certain is the necessity
to introduce {\em two\/} real scalar fields instead of one, since in the
noncommutative realm the U(1) subgroup of U(2) fails to decouple. What has been
missing is a guiding principle towards the ``correct'' field parametrization.

Since the sine-Gordon model can be obtained by dimensional reduction from
2+2 dimensional SDYM theory via a 2+1 dimensional integrable sigma model
\cite{ward}, and because the latter's noncommutative extension was shown to be
integrable in \cite{LP1}, it seems a good idea to contruct an integrable 
generalization of the sine-Gordon equation by starting from the linear system
of this integrable sigma model endowed with a time-space noncommutativity. 
This is the key strategy of this paper.
The reduction is performed on the equations of motion first, but it also works
at the level of the action, so giving directly the 1+1 dimensional action 
we are looking for. We interpret this success as an indication that the new
field parametrization proposed here is the proper one.

To be more precise, we are going to propose three different parametrizations,
by pairs of fields $(\phi_+,\phi_-)$, $(\rho,\vp)$ and $(h_1,h_2)$, all related
by nonlocal field redefinitions but all deriving from the compatibility
conditions of the underlying linear system \cite{LP1}. The first two appear 
in a ``Yang gauge''~\cite{Y} while the third one arises in a ``Leznov gauge''
\cite{L}. For either field pair in the Yang gauge,
the nontrivial compatibility condition reduces to a pair of 
``noncommutative sine-Gordon equations'' which in the commutative
limit degenerates to the standard sine-Gordon equation for 
$\sfrac12(\phi_+{+}\phi_-)$ or $\vp$, respectively, while 
$\sfrac12(\phi_+{-}\phi_-)$ or $\rho$ decouple as free bosons. 
The alternative Leznov formulation has the advantage
of producing two polynomial (actually, quadratic) equations of motion for 
$(h_1,h_2)$ but retains their coupling even in the commutative limit.

With the linear system comes a well-developed technology for generating
solitonic solutions to the equations of motion. Here, we shall employ the
dressing method \cite{dressing,faddeev} to explicitly outline the construction
of noncommutative sine-Gordon multi-solitons, directly in 1+1 dimensions as 
well as by reducing plane-wave solutions of the 2+1 dimensional integrable 
sigma model \cite{bieling}.
We completely analyze the one-soliton sector where we recover the standard
soliton solution as undeformed; noncommutativity becomes palpable only at the
multi-soliton level.

It was shown in \cite{LPS} that the tree-level $n$-point amplitudes of 
noncommutative 2+2 dimensional SDYM vanish for $n>3$,
consistent with the vanishing theorems for the $N{=}2$ string.
Therefore, we may expect nice properties of the S-matrix
to be inherited by our noncommutative sine-Gordon theory.
In fact, a direct evaluation of tree-level amplitudes reveals that,
in the Yang as well as the Leznov formulation,
the S-matrix is {\em causal\/} and no particle production occurs.

An overview of the paper is the following. In section~2 we review the 
basic construction of the 2+1 dimensional integrable sigma model 
of \cite{LP1} through a linear system, for the case of a noncommuting 
time coordinate. 
In section 3~we describe its dimensional reduction to the noncommutative 
integrable sine-Gordon model, both in the Yang and the Leznov formulation. 
Section~4 is devoted to the construction of solitonic solutions for our model,
by way of the iterative dressing approach.
The computation of scattering amplitudes is described in section~5. 
Finally, section~6 contains our conclusions and possible future directions.

\section{Noncommutative integrable sigma model in $2{+}1$ dimensions}

As has been known for some time, nonlinear sigma models in $2{+}1$ dimensions
may be Lorentz-invariant or integrable but not both~\cite{ward}.
Since the integrable variant serves as our starting point for the derivation
of the sine-Gordon model and its soliton solutions, we shall present its 
noncommutative extension \cite{LP1} in some detail in the present section.

\noindent
{\bf Noncommutative $\R^{2,1}$.\ }
Classical field theory on noncommutative spaces may be realized
by deforming the ordinary product of classical fields
(or their components) to the noncommutative star product
\begin{equation}
(f \star g)(x)\ =\ f(x)\,\exp\,\bigl\{ \frac{\im}{2}
{\ld{\partial}}_a \,\theta^{ab}\, {\rd{\partial}}_b \bigr\}\,g(x)\quad,
\end{equation}
with a constant antisymmetric tensor~$\t^{ab}$, where $a,b,\ldots=0,1,2$.
Specializing to $\R^{2,1}$, we shall use (real) coordinates $(x^a)=(t,x,y)$
in which the Minkowskian metric reads $(\eta_{ab})=\textrm{diag}(-1,+1,+1)$.
For later use we introduce the light-cone coordinates 
\begin{equation} \label{lightcone}
u\ :=\ \sfrac{1}{2}(t+y)\quad,\qquad
v\ :=\ \sfrac{1}{2}(t-y)\quad,\qquad
\pa_u\ =\ \pa_t+\pa_y\quad,\qquad
\pa_v\ =\ \pa_t-\pa_y \quad.
\end{equation} 
In view of the future reduction to $1{+}1$ dimensions, we choose the
coordinate~$x$ to remain commutative, so that the only non-vanishing
component of the noncommutativity tensor is 
\begin{equation}
\t^{ty}\ =\ -\t^{yt}\ =:\ \t\ >\ 0 \quad.
\end{equation}

\bigskip

\noindent
{\bf Linear system.\ }
Consider on noncommutative $\R^{2,1}$ 
the following pair of linear differential equations~\cite{LP1},
\begin{equation}\label{linsys}
(\z \pa_x -\pa_u)\Psi\ =\ A\star\Psi \qquad\textrm{and}\qquad
(\z \pa_v -\pa_x)\Psi\ =\ B\star\Psi \quad,
\end{equation}
where a spectral parameter~$\z\in\C P^1\cong S^2$ has been introduced.
The auxiliary field $\Psi$ takes values in U$(n)$ and depends on
$(t,x,y,\z)$ or, equivalently, on $(x,u,v,\z)$.
The $u(n)$ matrices $A$ and~$B$, in contrast, do not depend on~$\z$
but only on $(x,u,v)$.  
Given a solution~$\Psi$, they can be reconstructed via\footnote{
Inverses are understood with respect to the star product, 
i.e.~$\Psi^{-1}\star\Psi=\mbf{1}$.} 
\begin{equation} \label{ABfromPsi}
A \= \Psi\star(\pa_u-\z\pa_x)\Psi^{-1} \qquad\textrm{and}\qquad
B \= \Psi\star(\pa_x-\z\pa_v)\Psi^{-1} \quad.
\end{equation}
It should be noted that the equations (\ref{linsys}) 
are not of first order but actually of infinite order in derivatives, 
due to the star products involved.
In addition, the matrix $\Psi$ is subject to the following reality
condition~\cite{ward}:
\begin{equation}\label{real}
\mbf{1} \= \Psi(t,x,y,\z)\,\star\,[\Psi(t,x,y,\bar{\z})]^{\dagger} \quad,
\end{equation}
where `$\dagger$' is hermitian conjugation.
The compatibility conditions for the linear system~(\ref{linsys}) read
\begin{align}
\pa_x B -\pa_v A\ =\ 0 \quad ,\label{comp1} \\[4pt]
\pa_x A -\pa_u B -A\star B +B\star A\ =\ 0 \quad . \label{comp2}
\end{align}
By detailing the behavior of~$\Psi$ at small~$\z$ and at large~$\z$
we shall now ``solve'' these equations in two different ways, each one
leading to a single equation of motion for a particular field theory.

\bigskip

\noindent
{\bf Yang-type solution.\ }
We require that $\Psi$ is regular at $\z{=}0$~\cite{ivle1}, 
\begin{equation} \label{asymp1}
\Psi(t,x,y,\z\to0)\= \Phi^{-1}(t,x,y)\ +\ O(\z) \quad,
\end{equation}
which defines a U$(n)$-valued field $\Phi(t,x,y)$,
i.e.~$\ \Phi^\dagger=\Phi^{-1}$.
Therewith, $A$ and~$B$ are quickly reconstructed via
\begin{equation} \label{ABfromPsi2}
A\=\Psi\star\pa_u\Psi^{-1}\big|_{\z=0}
\=\Phi^{-1}\star\pa_u\Phi \qquad\textrm{and}\qquad
B\=\Psi\star\pa_x\Psi^{-1}\big|_{\z=0}
\=\Phi^{-1}\star\pa_x\Phi \quad.
\end{equation}
It is easy to see that compatibility equation (\ref{comp2}) is then automatic
while the remaining equation~(\ref{comp1}) turns into~\cite{LP1}
\begin{equation} \label{yangtype}
\pa_x\,(\Phi^{-1}\star\pa_x\Phi)-\pa_v\,(\Phi^{-1}\star\pa_u\Phi)\ =\ 0 \quad.
\end{equation}
This Yang-type equation~\cite{Y} can be rewritten as
\begin{equation} \label{yangtype2}
(\eta^{ab}+v_c\,\eps^{cab})\,\pa_a (\Phi^{-1}\star\pa_b \Phi)\ =\ 0\quad,
\end{equation}
where $\eps^{abc}$ is the alternating tensor with $\eps^{012}{=}1$ 
and $(v_c)=(0,1,0)$ is a fixed spacelike vector.
Clearly, this equation is not Lorentz-invariant but (deriving from a Lax pair)
it is integrable.

One can recognize (\ref{yangtype2}) as the field equation for
(a noncommutative generalization of) a WZW-like
modified U$(n)$ sigma model~\cite{ward,ioannidou} with the action\footnote{
which is obtainable by dimensional reduction from the Nair-Schiff
action~\cite{nair,moore} for SDYM in 2+2 dimensions}
\begin{equation} \label{Yaction}
\begin{aligned}
S_{\textrm{Y}}\ &=\ -\sfrac12\int\!\diff{t}\,\diff{x}\,\diff{y}\;\eta^{ab}\;
\tr\,\Bigl(\pa_a \Phi^{-1} \star\, \pa_b \Phi \Bigr) \\
&\quad\ -\sfrac13\int\!\diff{t}\,\diff{x}\,\diff{y} \int_0^1\!\diff{\la}\;
\widetilde{v}_{\r}\,\eps^{\r\m\n\s}\;\tr\,\Bigl(
\Pht^{-1}\star\,\pa_{\m}\Pht\,\star\,
\Pht^{-1}\star\,\pa_{\n}\Pht\,\star\,
\Pht^{-1}\star\,\pa_{\s}\Pht \Bigr) \quad,
\end{aligned}
\end{equation}
where Greek indices include the extra coordinate~$\la$,
and $\eps^{\r\m\n\s}$ denotes the totally antisymmetric tensor in~$\R^4$.
The field~$\Pht(t,x,y,\la)$ is an extension of~$\Phi(t,x,y)$, 
interpolating between
\begin{equation}
\Pht(t,x,y,0)\ =\ \textrm{const} \qquad\quad\textrm{and}\qquad\quad
\Pht(t,x,y,1)\ =\ \Phi(t,x,y) \quad,
\end{equation}
and `$\tr$' implies the trace over the U$(n)$ group space.
Finally, $(\widetilde{v}_{\r})=(v_c,0)$
is a constant vector in (extended) space-time. 

\bigskip

\noindent
{\bf Leznov-type solution.\ }
Finally, we also impose the asymptotic condition that 
$\ \lim_{\z\to\infty}\Psi=\Psi^0$ with some constant unitary 
(normalization) matrix~$\Psi^0$. The large~$\z$ behavior~\cite{ivle1}
\begin{equation} \label{asymp2}
\Psi(t,x,y,\z\to\infty)\= 
\bigl( \mbf{1}\ +\ \z^{-1}\Ups(t,x,y)\ +\ O(\z^{-2}) \bigr)\,\Psi^0
\end{equation}
then defines a $u(n)$-valued field $\Ups(t,x,y)$. 
Again this allows one to reconstruct $A$ and~$B$ through
\begin{equation} \label{ABfromPsi3}
A\=-\lim_{\z\to\infty} \bigl(\z\,\Psi\star\pa_x\Psi^{-1}\bigr) \=\pa_x\Ups
\qquad\textrm{and}\qquad
B\=-\lim_{\z\to\infty} \bigl(\z\,\Psi\star\pa_v\Psi^{-1}\bigr) \=\pa_v\Ups
\quad.
\end{equation}
In this parametrization, compatibility equation (\ref{comp1}) becomes an
identity but the second equation~(\ref{comp2}) turns into~\cite{LP1}
\begin{equation} \label{leznovtype}
\pa_x^2\Ups -\pa_u\pa_v\Ups -
\pa_x \Ups \star \pa_v \Ups + \pa_v \Ups \star \pa_x \Ups \ =\ 0 \quad.
\end{equation}

This Leznov-type equation~\cite{L} can also be obtained by extremizing
the action
\begin{equation} \label{Laction}
S_{\textrm{L}}\= \int\!\diff{t}\,\diff{x}\,\diff{y}\ \tr\,\Bigl\{
\sfrac12\,\eta^{ab}\,\pa_a \Ups \,\star\, \pa_b \Ups \ +\ 
\sfrac13\,\Ups \star 
\bigl( \pa_x \Ups\,\star\,\pa_v \Ups - \pa_v \Ups\,\star\,\pa_x \Ups \bigr) 
\Bigr\} \quad,
\end{equation}
which is merely cubic.

Obviously, the Leznov field $\Ups$ is related to the Yang field $\Phi$ 
through the non-local field redefinition
\begin{equation} \label{nonlocal}
\pa_x\Ups\=\Phi^{-1}\star\pa_u\Phi
\qquad\textrm{and}\qquad
\pa_v\Ups\=\Phi^{-1}\star\pa_x\Phi \quad.
\end{equation}
For each of the two fields $\Phi$ and~$\Ups$, one equation from the pair 
(\ref{comp1}, \ref{comp2}) represents the equation of motion,
while the other one is a direct consequence of the parametrization
(\ref{ABfromPsi2}) or~(\ref{ABfromPsi3}).

\section{Reduction to noncommutative sine-Gordon}

\noindent
{\bf Algebraic reduction ansatz.\ }
It is well known that the (commutative) sine-Gordon equation can be obtained
from the self-duality equations for SU(2) Yang-Mills upon appropriate
reduction from $2{+}2$ to $1{+}1$ dimensions. In this process the integrable 
sigma model of the previous section appears as an intermediate step in
$2{+}1$ dimensions, and so we may take its noncommutative extension as our 
departure point, after enlarging the group to U(2). 
In order to avoid cluttering the formulae we suppress the `$\star$' notation
for noncommutative multiplication from now on: all products are assumed to
be star products, and all functions are built on them, i.e.~$\e^{f(x)}$
stands for $\e_\star^{f(x)}$ and so on.

The dimensional reduction proceeds in two steps, firstly, a factorization
of the coordinate dependence and, secondly, an algebraic restriction of
the form of the U(2) matrices involved.
In the language of the linear system~(\ref{linsys}) the adequate ansatz
for the auxiliary field~$\Psi$ reads
\begin{equation} \label{ansatzPsi}
\Psi(t,x,y,\z) \= V(x)\,\psi(u,v,\z)\,V^\+(x)
\qquad\textrm{with}\qquad 
V(x) \= \Ecal\,\e^{\im\a\,x\,\s_1} \quad,
\end{equation}
where $\s_1=(\begin{smallmatrix} 0 & 1 \\ 1 & 0 \end{smallmatrix})$,
$\Ecal$ denotes some constant unitary matrix (to be specified later)
and $\a$ is a constant parameter. Under this factorization, 
the linear system~(\ref{linsys}) simplifies to\footnote{
The adjoint action means $\mathrm{ad}\s_1\,(\psi)=[\s_1,\psi]$.}
\begin{equation}
(\pa_u -\im\a\,\z\,\mathrm{ad}\s_1)\,\psi\=-a\,\psi \qquad\textrm{and}\qquad
(\z\pa_v -\im\a\,\mathrm{ad}\s_1)\,\psi\= b\,\psi 
\end{equation}
with $\ a=V^\+A\,V\ $ and $\ b=V^\+B\,V$.
Taking into account the asymptotic behavior (\ref{asymp1}, \ref{asymp2}), 
the ansatz~(\ref{ansatzPsi}) translates to the decompositions
\begin{align} \label{ansatzPhi}
\Phi(t,x,y) &\= V(x)\,g(u,v)\,V^\+(x)
\qquad\textrm{with}\qquad g(u,v) \in \textrm{U(2)}
\quad, \\[6pt] \label{ansatzUps}
\Ups(t,x,y) &\= V(x)\,\c(u,v)\,V^\+(x) 
\qquad\textrm{with}\qquad \c(u,v) \in u(2) \quad.
\end{align}
To aim for the sine-Gordon equation, one imposes certain algebraic 
constraints on $a$ and $b$ (and therefore on~$\psi$).
Their precise form, however, is not needed, as we are ultimately interested
only in $g$ or~$\c$. Therefore, we instead directly restrict $g(u,v)$
to the form
\begin{equation} \label{reducg}
g \= \Bigl(\begin{matrix} g_+ & 0 \\ 0 & g_- \end{matrix}\Bigr)
\= g_+ P_+ + g_- P_-
\qquad\textrm{with}\qquad g_+ \in \textrm{U(1)}_+ 
\quad\textrm{and}\quad    g_- \in \textrm{U(1)}_- 
\end{equation}
and with projectors $P_+=(\begin{smallmatrix}1&0\\0&0\end{smallmatrix})$
and $P_-=(\begin{smallmatrix}0&0\\0&1\end{smallmatrix})$.
This imbeds $g$ into a U(1)$\times$U(1) subgroup of~U(2).
Note that $g_+$ and $g_-$ do not commute, due to the implicit star product.
Invoking the field redefinition~(\ref{nonlocal}) we infer that the
corresponding reduction for~$\c(u,v)$ should be\footnote{
Complex conjugates of scalar functions are denoted with a dagger
to remind the reader of their noncommutativity.}
\begin{equation} \label{reduch}
\c \= \im \Bigl(\begin{matrix} 0 & h^\+ \\ h & 0 \end{matrix}\Bigr)
\qquad\textrm{with}\qquad h \in \C \quad,
\end{equation}
with the ``bridge relations''
\begin{equation} \label{nonlocal2}
\begin{aligned}
\a\,(h-h^\+) &\= - g_+^\+ \pa_u g_+ \= g_-^\+ \pa_u g_- \quad,\\[6pt]
\sfrac{1}{\a}\,\pa_v h &\= g_-^\+ g_+ - \mbf{1}
\qquad\textrm{and h.c.} \quad.
\end{aligned}
\end{equation}
In this way, the $u(2)$-matrix $\c$ is restricted to be off-diagonal.

We now investigate in turn the consequences of the ans\"atze 
(\ref{ansatzPhi}, \ref{reducg}) and (\ref{ansatzUps}, \ref{reduch}) 
for the equations of motion (\ref{yangtype}) and (\ref{leznovtype}),
respectively.

\bigskip

\noindent
{\bf Reduction of Yang-type equation.\ }
Let us insert the ansatz~(\ref{ansatzPhi}) into the
Yang-type equation of motion~(\ref{yangtype}).
After stripping off the $V$ factors one obtains
\begin{equation}
\pa_v (g^\+ \pa_u g) + \a^2 (\s_1 g^\+ \s_1 g - g^\+ \s_1 g \s_1) \= 0 \quad.
\end{equation}
Specializing with (\ref{reducg}) and employing the identities
$\ \s_1 P_\pm \s_1 = P_\mp\ $ we arrive at $\ Y_+P_+ + Y_-P_- =0$, 
with
\begin{equation} \label{Yg}
\begin{aligned}
Y_+ &\ \equiv\ 
\pa_v (g_+^\+ \pa_u g_+) + \a^2 (g_-^\+ g_+ - g_+^\+ g_-) \= 0 \quad, \\[6pt]
Y_- &\ \equiv\
\pa_v (g_-^\+ \pa_u g_-) + \a^2 (g_+^\+ g_- - g_-^\+ g_+) \= 0 \quad.
\end{aligned}
\end{equation}
Since the brackets multiplying~$\a^2$ are equal and opposite,
it is worthwhile to present the sum and the difference of the two equations:
\begin{equation} \label{Yg2}
\begin{aligned}
\pa_v\bigl( g_+^\+ \pa_u g_+ + g_-^\+ \pa_u g_- \bigr) &\= 0 \quad,\\[6pt]
\pa_v\bigl( g_+^\+ \pa_u g_+ - g_-^\+ \pa_u g_- \bigr) &\=
2\a^2 \bigl( g_+^\+ g_- - g_-^\+ g_+ \bigr) \quad.
\end{aligned}
\end{equation}

It is natural to introduce the angle fields $\p_\pm(u,v)$ via
\begin{equation} \label{para1}
g \= \e^{\frac{\im}{2}\p_+P_+}\,\e^{-\frac{\im}{2}\p_-P_-}
\qquad\Leftrightarrow\qquad
g_+ \= \e^{\frac{\im}{2}\p_+} 
\qquad\textrm{and}\qquad 
g_- \= \e^{-\frac{\im}{2}\p_-} \quad.
\end{equation}
In terms of these, the equations~(\ref{Yg2}) read
\begin{equation} \label{Yphi}
\begin{aligned}
\pa_v\bigl( \e^{-\frac{\im}{2}\p_+}\,\pa_u\e^{\frac{\im}{2}\p_+} +
            \e^{\frac{\im}{2}\p_-}\,\pa_u\e^{-\frac{\im}{2}\p_-} \bigr) &\=0
\quad,\\[6pt]
\pa_v\bigl( \e^{-\frac{\im}{2}\p_+}\,\pa_u\e^{\frac{\im}{2}\p_+} -
            \e^{\frac{\im}{2}\p_-}\,\pa_u\e^{-\frac{\im}{2}\p_-} \bigr) &\=
2\a^2\bigl( \e^{-\frac{\im}{2}\p_+}\e^{-\frac{\im}{2}\p_-} -
            \e^{\frac{\im}{2}\p_-}\e^{\frac{\im}{2}\p_+} \bigr) \quad.
\end{aligned}
\end{equation}
We propose to call these two equations 
``the noncommutative sine-Gordon equations''.
Besides their integrability (see later sections for consequences)
their form is quite convenient for studying the commutative limit.
When $\t\to0$, (\ref{Yphi}) simplifies to
\begin{equation} \label{Ycomm}
\pa_u\pa_v (\p_+{-}\p_-) \= 0 \qquad\textrm{and}\qquad
\pa_u\pa_v (\p_+{+}\p_-) \= -8\a^2\,\sin\sfrac12(\p_+{+}\p_-) \quad.
\end{equation}
Because the equations have decoupled we may choose 
\begin{equation}
\p_+ \= \p_- \ =:\ \p  \qquad\Leftrightarrow\qquad
g_+ \= g_-^\+ \qquad\Leftrightarrow\qquad
g \in \textrm{U(1)}_{\textrm{A}}
\end{equation}
and reproduce the familiar sine-Gordon equation
\begin{equation} \label{sG}
(\pa_t^2 -\pa_y^2)\,\p \= -4\a^2\,\sin\p \quad.
\end{equation}
One learns that in the commutative case the reduction is 
SU(2)$\to$U(1)$_{\textrm{A}}$ since the U(1)$_{\textrm{V}}$
degree of freedom $\p_+{-}\p_-$ is not needed.
The deformed situation, however, requires extending SU(2) to U(2),
and so it is imperative here to keep both U(1)s and work with 
{\it two\/} scalar fields.

Inspired by the commutative decoupling, one may choose another
distinguished parametrization of~$g$, namely
\begin{equation} \label{para2}
g_+ \= \e^{\frac{\im}{2}\r}\,\e^{\frac{\im}{2}\vp}
\qquad\textrm{and}\qquad
g_- \= \e^{\frac{\im}{2}\r}\,\e^{-\frac{\im}{2}\vp} \quad,
\end{equation}
which defines angles $\r(u,v)$ and $\vp(u,v)$ for the linear combinations 
$\textrm{U}(1)_{\textrm{V}}$ and $\textrm{U}(1)_{\textrm{A}}$, respectively.
Inserting this into (\ref{Yg}) one finds
\begin{equation} \label{Yrho}
\begin{aligned}
\pa_v\bigl( \e^{-\frac{\im}{2}\vp}\,\pa_u\e^{\frac{\im}{2}\vp} \bigr) + 
2\im\a^2\,\sin\vp &\= 
-\pa_v\bigl[\e^{-\frac{\im}{2}\vp}\e^{-\frac{\im}{2}\r}\,
(\pa_u \e^{\frac{\im}{2}\r}) \e^{\frac{\im}{2}\vp} \bigr] \quad,\\[6pt]
\pa_v\bigl( \e^{\frac{\im}{2}\vp}\,\pa_u\e^{-\frac{\im}{2}\vp} \bigr) -
2\im\a^2\,\sin\vp &\=
-\pa_v\bigl[\e^{\frac{\im}{2}\vp}\e^{-\frac{\im}{2}\r}\,
(\pa_u \e^{\frac{\im}{2}\r}) \e^{-\frac{\im}{2}\vp} \bigr] \quad.
\end{aligned}
\end{equation}
In the commutative limit, this system is easily decoupled to
\begin{equation} \label{sG2}
\pa_u \pa_v \r \=0 \qquad\textrm{and}\qquad
\pa_u \pa_v \vp + 4\a^2\,\sin\vp \= 0 \quad,
\end{equation}
revealing that $\ \r\to\frac12(\p_+{-}\p_-)\ $ and 
$\ \vp\to\frac12(\p_+{+}\p_-)=\p\ $ in this limit.

It is not difficult to write down an action for (\ref{Yg})
(and hence for (\ref{Yphi}) or (\ref{Yrho})). The relevant action may be
computed by reducing (\ref{Yaction}) with the help of (\ref{ansatzPhi})
and~(\ref{reducg}). The result takes the form
\begin{equation} \label{gaction}
S[g_+,g_-] \= S_{W}[g_+]\,+\,S_{W}[g_-]\,+\,\alpha^2\int\!\diff{t}\,\diff{y}\;
\bigl( g_+^{\dag} g_- + g_-^\dag g_+ - 2 \bigr) \quad,
\end{equation}
where $S_W$ is the abelian WZW action
\begin{equation} \label{WZWaction}
S_{W}[f]\ \equiv\ - \sfrac12\int\!\intd t\,\intd y\; \pa_v f^{-1}\; \pa_u f 
\,-\,\sfrac13\int\!\intd t\,\intd y\int_0^1\!\intd \la\;\eps^{\m\n\s}\,
\hat f^{-1}\pa_\m\hat f\;\hat f^{-1}\pa_\n\hat f\;\hat f^{-1}\pa_\s\hat f
\quad.
\end{equation}
Here $\hat{f}(\la)$ is a homotopy path satisfying the conditions
$\hat{f}(0) = 1$ and $\hat{f}(1) = f$.
Parametrizing $g_{\pm}$ as in (\ref{para2}) and using the Polyakov-Wiegmann
identity, the action for $\rho$ and $\varphi$ reads
\begin{equation} \label{rhophiaction}
\begin{aligned}
S[\rho,\varphi] &\= 2 S_{PC}\bigl[ \e^{\frac{\im}{2}\varphi}\bigr] \,+\,
2\a^2\int\!\intd t\,\intd y\;\bigl( \cos{\varphi} -1 \bigr)\,+\,
2 S_{W}\bigl[ \e^{\frac{\im}{2}\rho}\bigr] \\ 
&\qquad - \int\!\intd t\,\intd y\; 
\e^{-\frac{\im}{2}\rho}\,\pa_v \e^{\frac{\im}{2}\rho} 
\bigl( \e^{-\frac{\im}{2}\varphi}\,\pa_u \e^{\frac{\im}{2}\varphi}
     + \e^{\frac{\im}{2}\varphi}\,\pa_u \e^{-\frac{\im}{2}\varphi}\bigr)\quad,
\end{aligned}
\end{equation}
where
\begin{equation}
S_{PC}[f]\ \equiv\ -\sfrac12\int\!\intd t\,\intd y\;\pa_v f^{-1}\;\pa_u f\quad.
\end{equation}
In this parametrization the WZ term has apparently been shifted entirely to
the $\rho$ field while the cosine-type self-interaction remains for the
$\varphi$ field only. This fact has important consequences for the scattering
amplitudes.

It is well known \cite{witten, bosonization, bosonization2}
that in ordinary commutative geometry the bosonization of $N$ free massless 
fermions in the fundamental representation of SU($N$) gives rise to a WZW model
for a scalar field in SU($N$) plus a free scalar field associated with 
the U(1) invariance of the fermionic system.
In the noncommutative case the bosonization of a single massless Dirac fermion 
produces a noncommutative U(1) WZW model~\cite{MS}, which becomes free only 
in the commutative limit. Moreover, the U(1) subgroup of U($N$) does no 
longer decouple~\cite{matsubara}, so that $N$ noncommuting free massless 
fermions are related to a noncommutative WZW model for a scalar in U($N$).
On the other hand, giving a mass to the single Dirac fermion leads to a 
noncommutative cosine potential on the bosonized side~\cite{mass}.

In contrast, the noncommutative sine-Gordon model we propose in this paper 
is of a more general form.
The action~(\ref{gaction}) describes the propagation of a scalar field $g$ 
taking its value in U(1)$\times$U(1) $\subset$ U(2). Therefore, we expect it 
to be a bosonized version of two fermions in some representation of 
U(1)$\times$U(1). The absence of a WZ term for $\varphi$ and the lack
of a cosine-type self-interaction for~$\rho$ as well as the non-standard
interaction term make the precise identification non-trivial however.

\bigskip

\noindent
{\bf Reduction of Leznov-type equation.\ }
Alternatively, if we insert the ansatz~(\ref{ansatzUps}) into the 
Leznov-type equation of motion~(\ref{leznovtype}) we get
\begin{equation}
\pa_u\pa_v\c+2\a^2(\c-\s_1\c\s_1)+\im\a\bigl[[\s_1,\c],\pa_v\c\bigr]\=0\quad.
\end{equation}
Specializing with (\ref{reduch}) this takes the form $\ Z\s_-+Z^\+\s_+=0\ $
with $\ \s_-=(\begin{smallmatrix}0&0\\1&0\end{smallmatrix})\ $ and
$\ \s_+=(\begin{smallmatrix}0&1\\0&0\end{smallmatrix})$, where
\begin{equation} \label{Lh}
Z\ \equiv\ \pa_u\pa_v h + 2\a^2\,(h-h^\+)
+ \a\,\bigl\{ \pa_v h\,,\,h-h^\+ \bigr\} \=0 \quad.
\end{equation}
The decomposition
\begin{equation} \label{para3}
\c \= \im(h_1\s_1 + h_2\s_2) \qquad\Leftrightarrow\qquad h \= h_1 + \im h_2
\end{equation}
then yields
\begin{equation}
\begin{aligned} \label{Lh12}
\pa_u\pa_v h_1 -2\a\,\bigl\{ \pa_v h_2\,,\,h_2 \bigr\} &\=0 \quad, \\[6pt]
\pa_u\pa_v h_2 +4\a^2 h_2 +2\a\,\bigl\{ \pa_v h_1\,,\,h_2 \bigr\} &\=0 \quad.
\end{aligned}
\end{equation}
These two equations constitute an alternative description of the
noncommutative sine-Gordon model; they are classically equivalent to
the pair of~(\ref{Yg2}) or, to be more specific, to the pair of~(\ref{Yrho}). 
For the real fields the ``bridge relations''~(\ref{nonlocal2}) read
\begin{equation} \label{nonlocal3}
\begin{aligned}
& 2\im\a\,h_2 \= -\e^{-\frac{\im}{2}\vp}\e^{-\frac{\im}{2}\r}\,
\pa_u ( \e^{\frac{\im}{2}\r}\e^{\frac{\im}{2}\vp} )
\= \e^{\frac{\im}{2}\vp}\e^{-\frac{\im}{2}\r}\,
\pa_u ( \e^{\frac{\im}{2}\r}\e^{-\frac{\im}{2}\vp} ) \quad, \\[6pt]
& \qquad\qquad \sfrac{1}{\a}\pa_v h_1 \= \cos\vp-1 \qquad\textrm{and}\qquad
\sfrac{1}{\a}\pa_v h_2 \= \sin\vp \quad.
\end{aligned}
\end{equation}
One may ``solve'' one equation of~(\ref{Yrho}) by an appropriate field 
redefinition from~(\ref{nonlocal3}), which implies already one member 
of~(\ref{Lh12}). The second equation from~(\ref{Yrho}) then yields 
the remaining ``bridge relations'' in~(\ref{nonlocal3}) as well as 
the other member of~(\ref{Lh12}). This procedure works as well in the 
opposite direction, from~(\ref{Lh12}) to (\ref{Yrho}). 
The nonlocal duality between $(\vp,\r)$ and $(h_1,h_2)$ is simply a 
consequence of the equivalence between (\ref{yangtype}) and (\ref{leznovtype})
which in turn follows from our linear system~(\ref{linsys}).

The ``$h$~description'' has the advantage of being polynomial.
It is instructive to expose the action for the system~(\ref{Lh12}).
Either by inspection or by reducing the Leznov action~(\ref{Laction})
one obtains
\begin{equation}
S[h_1,h_2] \= \int\!\diff{t}\,\diff{y}\;
\Bigl\{ \pa_u h_1 \pa_v h_1 + \pa_u h_2 \pa_v h_2 
-4\a^2 h_2^2 -4\a\,h_2^2\,\pa_v h_1 \Bigr\} \quad.
\label{haction}
\end{equation}

\bigskip 

\noindent
{\bf Relation with other noncommutative generalizations of sine-Gordon.\ }
The noncommutative generalizations of the sine-Gordon model presented above
are expected to possess an infinite number of conservation laws, as they
originate from the reduction of an integrable model \cite{LPS2}. It is 
worthwhile to point out their relation to previously proposed noncommutative
sine-Gordon models which also feature an infinite number of local conserved 
currents.  

In \cite{GP} an alternative noncommutative version of the sine-Gordon model 
was proposed. Using the bicomplex approach the equations of motion were 
obtained as flatness conditions of a bidifferential calculus,\footnote{
This subsection switches to Euclidean space $\R^2$, where $\pa$ and
$\bar\pa$ are derivatives with respect to complex coordinates.}
\beq
\bar{\pa} ( G^{-1} \star \pa G) \= [ R\,,\, G^{-1} \star S\,G ]_{\star} \quad,
\label{eq}
\eeq
where  
\beq
R \= S \= 2\alpha\, \Bigl(\begin{matrix} 0 & 0 \\ 0 & 1 \end{matrix}\Bigr) 
\eeq
and $G$ is a suitable matrix in U(2) or, more generally, in complexified
U(2). In \cite{GP} the $G$ matrix was chosen as
\beq \label{sgdif}
G \= \e_{\star}^{\frac{\im}{2} \s_2 \phi} \= \biggl( \begin{matrix} 
\phantom{-}\cos_{\star}{\frac{\phi}{2}} & \ \sin_{\star}{\frac{\phi}{2}} 
\\[4pt]
-\sin_{\star}{\frac{\phi}{2}} & \ \cos_{\star}{\frac{\phi}{2}} 
\end{matrix} \biggr)
\eeq
with $\phi$ being a complex scalar field. This choice produces the 
noncommutative equations (all the products are $\star$-products)
\bea
&& 
\bar{\pa} \bigl( \e^{\frac{\im}{2} \phi}  \pa \e^{-\frac{\im}{2} \phi} 
+ \e^{-\frac{\im}{2} \phi}   \pa \e^{\frac{\im}{2} \phi} \bigr) 
~=~ 0 \quad,
\nonumber \\
&&
\bar{\pa} \bigl( \e^{-\frac{\im}{2} \phi}  \pa \e^{\frac{\im}{2} \phi} 
- \e^{\frac{\im}{2} \phi}   \pa \e^{-\frac{\im}{2} \phi} \bigr) 
~=~ 4\im\alpha^2 \sin{\phi} \quad.
\label{sg3}
\eea
As shown in \cite{GMPT} these equations (or a linear combination of them) 
can be obtained as a dimensional reduction of the equations of motion
for noncommutative U(2) SDYM in 2+2 dimensions.

The equations (\ref{sg3}) can also be derived from an action which consists 
of the sum of two WZW actions augmented by a cosine potential,
\beq
S[f,\bar f]\=S[f]+S[\bar f]  \qquad\text{with}\qquad 
S[f] \ \equiv\ S_W[f] -
\alpha^2 \int\!\diff{t}\,\diff{y}\; \bigl( f^2+ f^{-2} -2 \bigr) \quad,
\label{sg4}
\eeq
with $S_W[f]$ given in (\ref{WZWaction}) for $f\equiv\e^{\frac{\im}{2}\phi}$
in complexified U(1). 
However, this action cannot be obtained from the SDYM action in 2+2 dimensions
by performing the same field parametrization which led to (\ref{sg3}). 

Comparing the actions (\ref{gaction}) and (\ref{sg4}) 
and considering $f$ and $\bar{f}$ as independent U(1) group valued fields
we are tempted to formally identify $f \equiv g_+$ and $\bar{f} \equiv g_-$. 
Doing this, we immediately realize that the two models differ in
their interaction term which generalizes the cosine potential. 
While in (\ref{sg4}) the fields $f$ and $\bar{f}$ show only self-interaction,
the fields $g_+$ and $g_-$ in (\ref{gaction}) interact with each other. 
As we will see in section 5 this makes a big difference when evaluating the
S-matrix elements.

We close this section by observing that the equations of motion (\ref{Yphi})
can also be obtained directly in two dimensions by using the bicomplex
approach described in \cite{GP}. In fact, if instead of (\ref{sgdif}) we choose
\beq
G \= 
\biggl( \begin{matrix} 
\e^{\frac{\im}{2}\phi_+} + \e^{-\frac{\im}{2}\phi_-}
& \ -\im \e^{\frac{\im}{2}\phi_+} +\im \e^{-\frac{\im}{2}\phi_-} \\[4pt]
\im \e^{\frac{\im}{2}\phi_+} -\im \e^{-\frac{\im}{2}\phi_-}
& \phantom{-}\ \e^{\frac{\im}{2}\phi_+} + \e^{-\frac{\im}{2}\phi_-} 
\end{matrix} \biggr)
\eeq
it is easy to prove that (\ref{eq}) yields exactly the set of 
equations (\ref{Yphi}). Therefore, by exploiting the results in \cite{GP}
it should be straightforward to construct the first nontrivial conserved 
currents for the present model.

\section{Noncommutative solitons}

\noindent
{\bf Dressing approach in 2+1 dimensions.\ }
The existence of the linear system allows for powerful methods to
systematically construct explicit solutions for $\Psi$ and hence
for $\Phi^\+=\Psi|_{\z=0}$ or $\Ups$. 
For our purposes the so-called dressing method
\cite{dressing,faddeev}
proves to be most practical, and so we shall first present it here for
our linear system~(\ref{linsys}), before reducing the results to
solitonic solutions of the noncommutative sine-Gordon equations.

The central idea is to demand analyticity in the spectral parameter~$\z$
for the linear system~(\ref{linsys}), which strongly restricts the possible
form of~$\Psi$. The most elegant way to exploit this constraint starts from
the observation that the left hand sides of the differential relations
(D):=(\ref{ABfromPsi}) as well as the reality condition (R):=(\ref{real})
do not depend on~$\z$ while their right hand sides
are expected to be nontrivial functions of~$\z$ (except for the
trivial case $\Psi=\Psi^0$). More specifically, $\C P^1$ being compact,
the matrix function~$\Psi(\z)$ cannot be holomorphic everywhere but must
possess some poles, and hence the right hand sides of (D) and
(R) should display these (and complex conjugate) poles as well.
The resolution of this conundrum demands that the residues of the 
right hand sides at any would-be pole in~$\z$ have to vanish. 
We are now going to evaluate these conditions.

The dressing method builds a solution $\Psi_N(t,x,y,\z)$ featuring
$N$~simple poles at positions $\m_1,\m_2,\ldots,\m_N$ by left-multiplying 
an $(N{-}1)$-pole solution $\Psi_{N-1}(t,x,y,\z)$ with a single-pole factor 
of the form $\ \bigl(1+\frac{\m_N{-}\mb_N}{\z{-}\m_N}P_N(t,x,y)\bigr)$, 
where the $n{\times}n$ matrix function $P_N$ is yet to be determined. 
In addition, we are free to right-multiply $\Psi_{N-1}(t,x,y,\z)$ with some
constant unitary matrix~$\Psh^0_N$.  Starting from $\Psi_0=\mbf{1}$, 
the iteration $\ \Psi_0\mapsto\Psi_1\mapsto\ldots\mapsto\Psi_N\ $
yields a multiplicative ansatz for $\Psi_N$ which, via partial fraction 
decomposition, may be rewritten in an additive form (as a sum of simple pole
terms). Let us trace this iterative procedure constructively.

In accord with the outline above, the one-pole ansatz must read
($\Psh^0_1=:\Psi^0_1$)
\begin{equation} \label{Psione}
\Psi_1 \= \Bigl(\mbf{1}\,+\,\frac{\m_1-\mb_1}{\z-\m_1}\,P_1\Bigr)\,\Psi^0_1
\= \Bigl(\mbf{1} \,+\,\frac{\Lambda_{11}S_1^\+}{\z-\m_1}\Bigr)\,\Psi^0_1
\end{equation}
with some $n{\times}r_1$ matrix functions $\Lambda_{11}$ and~$S_1$
for some $1{\le}r_1{<}n$. The normalization matrix~$\Psi^0_1$ is 
constant and unitary. It is quickly checked that
\begin{equation} \label{R1}
\res_{\z=\mb_1} (R) =0 \qquad\Longrightarrow\qquad
P_1^\+ \= P_1 \= P_1^2 \qquad\Longrightarrow\qquad
P_1 \= T_1\,(T_1^\+ T_1)^{-1} T_1^\+ \quad,
\end{equation}
meaning that $P_1$ is a rank~$r_1$ projector built from an $n{\times}r_1$ 
matrix function~$T_1$. The columns of~$T_1$ span the image of~$P_1$ and 
obey $P_1T_1=T_1$. When using the second parametrization of~$\Psi_1$
in~(\ref{Psione}) one finds that
\begin{equation} \label{R2}
\res_{\z=\mb_1} (R) =0 \qquad\Longrightarrow\qquad
(\mbf{1}-P_1)\,S_1\Lambda_{11}^\+ \=0 \qquad\Longrightarrow\qquad
T_1 \= S_1 \qquad\qquad\qquad{}
\end{equation}
modulo a freedom of normalization.
Finally, the differential relations yield
\begin{equation} \label{D1}
\res_{\z=\mb_1} (D) =0 \qquad\Longrightarrow\qquad
(\mbf{1}-P_1)\,\Lb_1\,(S_1\Lambda_{11}^\+) \=0 \qquad\Longrightarrow\qquad
\Lb_1\,S_1 \= S_1\,\Ga_1^{A,B}
\end{equation}
for some $r_1{\times}r_1$ matrices~$\Ga_1^A$ and $\Ga_1^B$, 
after having defined
\begin{equation}
\bar{L}_i^A\ :=\ \pa_u-\mb_i\pa_x \qquad\textrm{and}\qquad 
\bar{L}_i^B\ :=\ \m_i(\pa_x-\mb_i\pa_v) \qquad\textrm{for}\quad
i=1,2,\ldots,N \quad.
\end{equation}
Because the $\Lb_i$ are linear differential operators it is easy to
write down the general solution for~(\ref{D1}): 
Introduce ``co-moving coordinates''
\begin{equation} \label{comoving}
w_i \ :=\ x + \mb_i u + \mb_i^{-1} v \qquad\Longrightarrow\qquad
\bar{w}_i \= x + \m_i u + \m_i^{-1} v \qquad\textrm{for}\quad
i=1,2,\ldots,N 
\end{equation}
so that on functions of $(w_i,\bar{w}_i)$ alone the $\Lb_i$ act as
\begin{equation}
\bar{L}^A_i \= \bar{L}^B_i \= (\m_i{-}\mb_i)\frac{\pa}{\pa\bar{w}_i} \quad.
\end{equation}
Hence, (\ref{D1}) is solved by
\begin{equation}
S_1(t,x,y) \= \Sh_1(w_1)\,\e^{\bar{w}_1 \Ga_1 /(\m_1-\mb_1)} 
\qquad\textrm{ 
for any $w_1$-holomorphic $n{\times}r_1$ matrix function $\Sh_1$}
\end{equation}
and $\Ga_1^A=\Ga_1^B=:\Ga_1$.
Appearing to the right of~$\Sh_1$, the exponential factor is seen to drop out
in the formation of~$P_1$ via (\ref{R1}) and~(\ref{R2}). Thus, no generality
is lost by taking $\Ga_1=0$. We learn that any $w_1$-holomorphic 
$n{\times}r_1$ matrix $T_1$ is admissible to build a projector~$P_1$
which then yields a solution $\Psi_1$ (and thus~$\Phi$) via~(\ref{Psione}).
Note that $\Lambda_{11}$ need not be determined seperately but follows from 
our above result.
It is not necessary to also consider the residues at $\z{=}\m_1$ since
their vanishing leads merely to the hermitian conjugated conditions.

Let us proceed to the two-pole situation. The dressing ansatz takes the form
($\Psi^0_1\Psh^0_2=:\Psi^0_2$)
\begin{equation} \label{Psitwo}
\Psi_2 \= \Bigl(\mbf{1} \,+\,\frac{\m_2-\mb_2}{\z-\m_2}\,P_2\Bigr)
\Bigl(\mbf{1} \,+\,\frac{\m_1-\mb_1}{\z-\m_1}\,P_1\Bigr) \,\Psi^0_2
\= \Bigl(\mbf{1} \,+\,\frac{\Lambda_{21}S_1^\+}{\z-\m_1}
\,+\,\frac{\Lambda_{22}S_2^\+}{\z-\m_2}\Bigr) \,\Psi^0_2 \quad,
\end{equation}
where $P_2$ and $S_2$ are to be determined but $P_1$ and $S_1$ can be copied 
from above. Indeed,
inspecting the residues of (R) and (D) at $\z=\mb_1$ simply confirms that
\begin{equation}
P_1 \= T_1\,(T_1^\+ T_1)^{-1} T_1^\+ \qquad\textrm{and}\qquad 
T_1\=S_1 \qquad\textrm{with}\qquad 
S_1 \= \Sh_1(w_1)
\end{equation}
is just carried over from the one-pole solution.
Relations for $P_2$ and $S_2$ arise from 
\begin{align}
\res_{\z=\mb_2} (R) =0 &\quad\Longrightarrow\quad
(\mbf{1}{-}P_2)\,P_2\=0\quad\Longrightarrow\quad
P_2 \= T_2\,(T_2^\+ T_2)^{-1} T_2^\+ \ ,\\[6pt]
\res_{\z=\mb_2} (R) =0 &\quad\Longrightarrow\quad
\Psi_2(\mb_2)\,S_2\Lambda_{22}^\+ \= 
(\mbf{1}{-}P_2)(1-\sfrac{\m_1-\mb_1}{\m_1-\mb_2}P_1)\,S_2\Lambda_{22}^\+\=0\ ,
\label{TnotS}
\end{align}
where the first equation makes use of the multiplicative form of the
ansatz~(\ref{Psitwo}) while the second one exploits the additive version.
We conclude that $P_2$ is again a hermitian projector (of some rank~$r_2$)
and thus built from an $n{\times}r_2$ matrix function~$T_2$. Furthermore,
(\ref{TnotS}) reveals that $T_2$ cannot be identified with $S_2$ this time,
but we rather have
\begin{equation} \label{T2fromS2}
T_2 \= \Bigl(1-\frac{\m_1{-}\mb_1}{\m_1{-}\mb_2}\,P_1\Bigr)\,S_2
\end{equation}
instead. Finally, we consider
\begin{equation}
\res_{\z=\mb_2} (D) =0 \qquad\Longrightarrow\qquad
\Psi_2(\mb_2)\,\Lb_2\,(S_2\Lambda_{22}^\+) \=0 \qquad\Longrightarrow\qquad
\Lb_2\,S_2 \= S_2\,\Ga_2^{A,B}
\end{equation}
which is solved by
\begin{equation}
S_2(t,x,y) \= \Sh_2(w_2)\,\e^{\bar{w}_2 \Ga_2 /(\m_2-\mb_2)} 
\qquad\textrm{ 
for any $w_2$-holomorphic $n{\times}r_2$ matrix function $\Sh_2$}
\end{equation}
and $\Ga_2^A=\Ga_2^B=:\Ga_2$. Once more, we are entitled to put $\Ga_2=0$.
Hence, the second pole factor in (\ref{Psitwo})
is constructed in the same way as the first one, except for the small
complication~(\ref{T2fromS2}). Again, $\Lambda_{21}$ and $\Lambda_{22}$ can
be read off the result if needed.

It is now clear how the iteration continues. 
After $N$ steps the final result reads
\begin{equation}
\Psi_N \= \biggl\{ \prod_{\ell=0}^{N-1} \Bigl(\mbf{1} \,+\, 
\frac{\m_{N-\ell}-\mb_{N-\ell}}{\z-\m_{N-\ell}}\,P_{N-\ell} \Bigr)\biggr\}\,
\Psi^0_N
\=\biggl\{\mbf{1}\,+\,\sum_{i=1}^N\frac{\Lambda_{Ni}S_i^\+}{\z-\m_i}\biggr\}
\,\Psi^0_N \quad,
\end{equation}
featuring hermitian rank $r_i$ projectors~$P_i$ at $i=1,2,\ldots,N$, via
\begin{equation}
P_i \= T_i\,(T_i^\+ T_i)^{-1} T_i^\+ \qquad\textrm{with}\qquad
T_i \= \biggl\{ \prod_{\ell=1}^{i-1} \Bigl(\mbf{1} \,-\,
\frac{\m_{i-\ell}-\mb_{i-\ell}}{\m_{i-\ell}-\mb_i}\,P_{i-\ell}\Bigr)\biggl\}
\,S_i
\quad,
\end{equation}
where
\begin{equation}
S_i(t,x,y) \= \Sh_i(w_i)
\end{equation}
for arbitrary $w_i$-holomorphic $n{\times}r_i$ matrix functions $\Sh_i(w_i)$.
The corresponding classical Yang and Leznov fields are
\begin{align} \label{PhiN}
\Phi_N &\= \Psi_N^\+(\z{=}0) \= 
{\Psi^0_N}^\+\,\prod_{i=1}^N \bigl( \mbf{1}-\r_i\,P_i \bigr)
\qquad\textrm{with}\qquad \r_i \= 1-\frac{\m_i}{\mb_i} \quad, \\[6pt]
\Ups_N &\=\lim_{\z\to\infty}\z\,\bigl(\Psi_N(\z)\,{\Psi^0_N}^\+-\mbf{1}\bigr)
\= \sum_{i=1}^N (\m_i{-}\mb_i)\,P_i \quad.
\end{align}
The solution space constructed here is parametrized (slightly redundantly) 
by the set $\{\Sh_i\}_1^N$ of matrix-valued holomorphic functions and the 
pole positions~$\m_i$.
The so-constructed classical configurations have solitonic character
(meaning finite energy) when all these functions are algebraic.

The dressing technique as presented above is well known in the commutative
theory; novel is only the realization that it carries over verbatim to the
noncommutative situation by simply understanding all products as star
products (and likewise inverses, exponentials, etc.). Of course, it may be
technically difficult to $\star$-invert some matrix, but one may always
fall back on an expansion in powers of~$\t$.

\bigskip 

\noindent
{\bf Solitons of the noncommutative sine-Gordon theory.\ }
We should now be able to generate $N$-soliton solutions to the noncommutative
sine-Gordon equations, say~(\ref{Yrho}), by applying the reduction from 
$2{+}1$ to $1{+}1$ dimensions (see previous section) to the above strategy
for the group~U(2), i.e.~putting $n{=}2$. 
In order to find nontrivial solutions, we specify the constant matrix~$\Ecal$
in the ansatz~(\ref{ansatzPsi}) for~$\Psi$ as
\begin{equation}
\Ecal \= \e^{-\im\frac{\pi}{4}\s_2} \= \sfrac{1}{\sqrt{2}}
\Bigl(\begin{matrix} 1 & -1 \\ 1 & \phantom{-} 1 \end{matrix}\Bigr)
\end{equation}
which obeys the relations
$\ \Ecal\s_3=\s_1\,\Ecal\ $ and $\ \Ecal\s_1=-\s_3\,\Ecal$.
Pushing $\Ecal$ beyond~$V$ we can write
\begin{equation} \label{Wdef}
\Phi(t,x,y) \= W(x)\,\gt(u,v)\,W^\+(x)
\qquad\textrm{with}\qquad W(x) \= \e^{-\im\a\,x\,\s_3}
\end{equation}
and
\begin{equation} \label{grot}
\gt(u,v) \= \Ecal\,g(u,v)\,\Ecal^\+ \= \Ecal\,\biggl( \begin{matrix} 
g_+ & \ 0 \\[4pt] 0 & \ g_- \end{matrix}\biggr)\,\Ecal^\+
\= \sfrac12 \, \biggl(\begin{matrix}
g_+{+}g_- & \ \ g_+{-}g_- \\[4pt] g_+{-}g_- & \ \ g_+{+}g_- 
\end{matrix}\biggr) \quad.
\end{equation}

With hindsight from the commutative case~\cite{faddeev} we choose
\begin{equation}
\Psh^0_i \= \s_3 \quad\forall i \qquad\Longleftrightarrow\qquad 
\Psi^0_N \=\s_3^N
\end{equation}
(which commutes with $W$)
and restrict the poles of~$\Psi$ to the imaginary axis, 
$\m_i=\im p_i\ $ with $\ p_i\in\R$.
Therewith, the co-moving coordinates~(\ref{comoving}) become
\begin{equation}
w_i \= x - \im (p_i\,u - p_i^{-1} v) \ =:\ x - \im\h_i(u,v) \quad,
\end{equation}
defining $\h_i$ as real linear functions of the light-cone coordinates.
Consequentially, from~(\ref{PhiN}) we get $\ \r_i=2\ $ and find that
\begin{equation}
\gt_N(u,v) \= \s_3^N\,\prod_{i=1}^N \bigl( \mbf{1}-2\,\Pt_i(u,v) \bigr)
\qquad\textrm{with}\qquad P_i \= W\,\Pt_i\,W^\+ \quad.
\end{equation}
Repeating the analysis of the previous subsection, 
one is again led to construct hermitian projectors
\begin{equation}
\Pt_i \= \Tt_i\,(\Tt_i^\+ \Tt_i)^{-1} \Tt_i^\+ \qquad\textrm{with}\qquad
\Tt_i \= \prod_{\ell=1}^{i-1} \Bigl(\mbf{1} \,-\,
\frac{2\,p_{i-\ell}}{p_{i-\ell}+p_i}\,\Pt_{i-\ell}\Bigr)\,\St_i \quad,
\end{equation}
where $2{\times}1$ matrix functions $\St_i(u,v)$ are subject to
\begin{equation} \label{Dred}
\Lbt_i\,\St_i \= \St_i\,\widetilde{\Ga}_i
\qquad\textrm{for}\quad  i=1,2,\ldots,N
\end{equation}
and some numbers~$\widetilde\Ga_i$ (note that now rank $r_i{=}1$) 
which again we can put to zero.
On functions of the reduced co-moving coordinates~$\h_i$ alone, 
\begin{equation}
\Lbt_i \= W^\+ \Lb_i W \=
(\m_i{-}\mb_i)\, W^\+ \frac{\pa}{\pa\bar{w}_i} W \= 
p_i\,\Bigl( \frac{\pa}{\pa\h_i} + \a \,\s_3 \Bigr)
\end{equation}
so that (\ref{Dred}) is solved by
\begin{equation}
\St_i(u,v) \= \widehat{\St}_i(\h_i) \= 
\biggl(\begin{matrix}\g_{i1}\e^{-\a\,\h_i}\\[4pt] 
                     \im\g_{i2}\e^{+\a\,\h_i}\end{matrix}\biggr)
\=\e^{-\a\,\h_i\s_3}\,
\biggl(\begin{matrix} \g_{i1} \\[4pt] \im\g_{i2} \end{matrix}\biggr)
\qquad\textrm{with}\quad \g_{i1}, \g_{i2} \in\C \quad.
\end{equation}
Furthermore, it is useful to rewrite
\begin{equation}
\g_{i1}\g_{i2} =: \la_i^2 \quad\textrm{and}\quad \g_{i2}/\g_{i1} =: \g_i^2
\qquad\Longleftrightarrow\qquad
\biggl(\begin{matrix} \g_{i1} \\[4pt] \im\g_{i2} \end{matrix}\biggr)\=\la_i\,
\biggl(\begin{matrix} \g_i^{-1} \\[4pt] \im\g_i \end{matrix}\biggr)
\end{equation}
because then $|\g_i|$ may be absorbed into $\h_i$ by shifting 
$\a\h_i\mapsto\a\h_i+\ln|\g_i|$. The multipliers~$\la_i$ drop out in the
computation of~$\Pt_i$.
Finally, to make contact with the form~(\ref{grot}) we restrict the
constants $\g_i$ to be real.

Let us check the one-soliton solution, i.e.~put $N{=}1$.
Suppressing the indices momentarily, 
absorbing $\g$ into~$\h$ and dropping~$\la$, we infer that
\begin{equation} \label{onesol}
\Tt\=\biggl(\begin{matrix} \e^{-\a\h} \\[4pt] \im\e^{\a\h} \end{matrix}\biggr)
\quad\Longrightarrow\quad
\Pt\=\frac{1}{2\,\ch 2\a\h} \biggl(\begin{matrix}
\e^{-2\a\h} & -\im \\[4pt] \im & \e^{+2\a\h} \end{matrix}\biggr)
\quad\Longrightarrow\quad
\gt\= \Biggl(\begin{matrix}
\th 2\a\h & \ \frac{\im}{\ch 2\a\h}\\[6pt] \frac{\im}{\ch 2\a\h} & \ \th 2\a\h 
\end{matrix}\Biggr)
\end{equation}
which has $\det\,\gt=1$. 
Since here the entire coordinate dependence comes in the single 
combination~$\h(u,v)$, all star products trivialize and the one-soliton 
configuration coincides with the commutative one. Hence, the field~$\r$ 
drops out, $\gt\in$ SU(2), and we find, comparing (\ref{onesol}) with 
(\ref{grot}), that
\begin{equation}
\sfrac12(g_+{+}g_-) \= \cos\sfrac{\vp}{2} \= \th 2\a\h
\qquad\textrm{and}\qquad
\sfrac{1}{2\im}(g_+{-}g_-) \= \sin\sfrac{\vp}{2} \= \sfrac{1}{\ch 2\a\h}
\end{equation}
which implies
\begin{equation}
\tan\sfrac{\vp}{4} \= \e^{-2\a\h} \qquad\Longrightarrow\qquad
\vp \= 4\,\arctan \e^{-2\a\h} \= -2\,\arcsin (\th 2\a\h) \quad,
\end{equation}
reproducing the well known sine-Gordon soliton with mass $\ m=2\a$. 
Its moduli parameters are the velocity $\ \n=\frac{1-p^2}{1+p^2}\ $ 
and the center of inertia $\ y_0=\frac{1}{\a}\sqrt{1{-}\n^2}\ln|\g|\ $ 
at zero time~\cite{faddeev}. In passing we note that in the 
``$h$~description'' the soliton solution takes the form
\begin{equation}
h_1 \= p\,\th 2\a\h \qquad\textrm{and}\qquad 
h_2 \= \sfrac{p}{\ch 2\a\h} \qquad\Longrightarrow\qquad
h \= p\,\th(\a\h{+}\sfrac{\im\pi}{4}) \= p\,\e^{\frac{\im}{2}\vp} \quad.
\end{equation}

Noncommutativity becomes relevant for multi-solitons. 
At $N{=}2$, for instance, one has
\begin{equation}
\begin{aligned}
& \gt_2 \= (1-2\Pt_1)\,(1-2\Pt_2) \qquad\textrm{with}\quad
\Pt_1 = \Pt \quad\textrm{from (\ref{onesol})}\quad\textrm{and}\quad
\Pt_2 \= \Tt_2\,(\Tt_2^\+ \Tt_2)^{-1} \Tt_2^\+ \\[6pt]
& \textrm{where}\qquad
\Tt_2 \= \bigl(\mbf{1} - \sfrac{2 p_1}{p_1+p_2} \Pt_1\bigr)\,\widehat{\St}_2
\qquad\textrm{and}\qquad \widehat{\St}_2 \= \e^{-\a\,\h_2\s_3}\,
\bigl(\begin{smallmatrix} \g_2^{-1} \\ \im\g_2 \end{smallmatrix}\bigr)
\quad\ \textrm{with}\quad \g_2\in\R \quad.
\end{aligned}
\end{equation}
We refrain from writing down the lengthy explicit expression for~$\gt_2$
in terms of the noncommuting coordinates $\h_1$ and~$\h_2$, 
but one cannot expect to find a unit (star-)determinant for~$\gt_2$
except in the commutative limit. This underscores the necessity of extending
the matrices to~U(2) and the inclusion of a nontrivial $\r$ 
at the multi-soliton level.

It is not surprising that the just-constructed noncommutative sine-Gordon 
solitons themselves descend directly from BPS solutions of the $2{+}1$
dimensional integrable sigma model. Indeed, putting back the $x$~dependence
via~(\ref{Wdef}), the $2{+}1$ dimensional projectors~$P_i$ are built from
$2{\times}1$ matrices
\begin{equation}
S_i \= W(x)\,\Sh_i(\h_i) 
\= \e^{-\im\a\,w_i\s_3}\, 
\biggl(\begin{matrix} \g_i^{-1} \\[4pt] \im\g_i \end{matrix}\biggr) \= 
\biggl(\begin{matrix} 1 \\[4pt] \im\g_i^2\e^{2\im\a\,w_i}\end{matrix}\biggr)\,
\g_i^{-1}\e^{-\im\a\,w_i} \quad.
\end{equation}
In the last expression the right factor drops out on the computation of
projectors; the remaining column vector agrees with the standard conventions
\cite{ward,LP1,faddeev,bieling}. Reassuringly, the coordinate dependence has 
combined into~$w_i$. The ensueing $2{+}1$ dimensional configurations~$\Phi_N$
are nothing but noncommutative multi-plane-waves the simplest examples of 
which were already investigated in~\cite{bieling}.

\section{Tree amplitudes}

\noindent
In this section we compute tree-level amplitudes for the noncommutative 
generalization of the sine-Gordon model proposed in section 3, 
both in the Yang and the Leznov formulation.
In commutative geometry the sine-Gordon S-matrix factorizes in two-particle
processes and no particle production occurs, as a consequence of 
the existence of an infinite number of conservation laws. 
In the noncommutative case it is interesting to investigate whether 
the presence of an infinite number of conserved currents is still
sufficient to guarantee the integrability of the system in the sense 
of having a factorized S-matrix.

A previous noncommutative version of the sine-Gordon model with an infinite 
set of conserved currents was proposed in \cite{GP}, and its S-matrix was 
studied in \cite{GMPT}. 
Despite the existence of an infinite chain of conservation laws, 
it turned out that particle production occurs in this model and that
the S-matrix is neither factorized nor causal.\footnote{
Acausal behaviour in noncommutative field theory was first observed in 
\cite{seiberg} and shown to be related to time-space noncommutativity.}
As already stressed in section 3, the noncommutative generalization of the 
sine-Gordon model we propose in this paper differs from the one studied 
in \cite{GP} in the generalization of the cosine potential. Therefore, both 
theories describe the dynamics of two real scalar fields, but 
the structure of the interaction terms between the two fields is different. 
We then expect the scattering amplitudes of the present
theory to behave differently from those of the previous one. To this end 
we will compute the amplitudes corresponding to $2\to 2$ processes for 
the fields $\rho$ and $\vp$ in the $g$-model (Yang formulation) as well as 
for the fields $h_1$ and $h_2$ in the $h$-model (Leznov formulation). 
In the $g$-model we will also compute $2\to 4$ and $3\to 3$ amplitudes 
for the massive field~$\vp$. In both models the S-matrix will turn out to be 
{\em factorized\/} and {\em causal\/} in spite of their time-space 
noncommutativity. 

\bigskip

\noindent
{\bf Amplitudes in the ``$g$-model''. Feynman rules.\ }
We parametrize the $g$-model with $(\rho, \varphi)$ as in (\ref{rhophiaction})
since in this parametrization the mass matrix turns out to be diagonal,
with zero mass for $\rho$ and $m{=}2\alpha$ for $\varphi$.
Expanding the action (\ref{rhophiaction}) up to the fourth order in the
fields, we read off the following Feynman rules: 
\begin{itemize}
\item{The propagators 
\bea
\parbox{2cm}{\includegraphics[width=1.9cm]{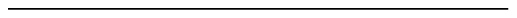}}
&\ \equiv\ &\langle \varphi\varphi\rangle\=\frac{2\im}{k^2-4\a^2}\quad,\\[4pt]
\parbox{2cm}{\includegraphics[width=1.9cm]{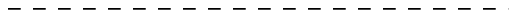}}
&\ \equiv\ &\langle \rho\,\rho \rangle\=\frac{2\im}{k^2}\quad.
\ena}
\item{The vertices 
(including a factor of ``i'' from the expansion of $\e^{\im S}$)
\bea
\parbox{2cm}{\includegraphics[width=1.9cm]{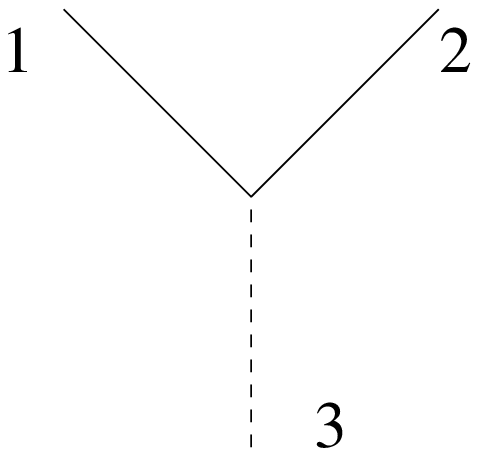}}
&\=&-\frac{1}{2^3}(k_{2}^2-k_1^2-2k_1\wedge k_2) F(k_1,k_2,k_3)\quad,\\
\parbox{2cm}{\includegraphics[width=1.9cm]{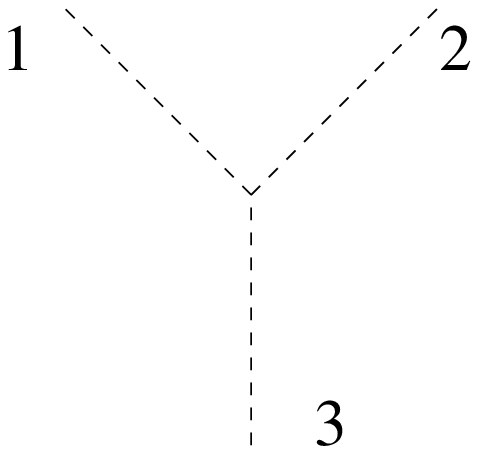}}
&\=&\frac{1}{2\cdot 3!}\;k_{1}\wedge k_2\; F(k_1,k_2,k_3)\quad, 
\ena}
\bea
\parbox{2cm}{\includegraphics[width=1.9cm]{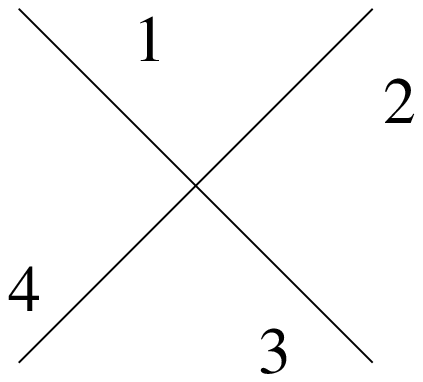}}
&\=&\Bigl[-\frac{\im}{2^3\cdot4!}\,(k_1^2+3k_1\cdot k_3)+
\frac{2\im\a^2}{4!}\Bigr]F(k_1,k_2,k_3,k_4)\quad,\\
\parbox{2cm}{\includegraphics[width=1.9cm]{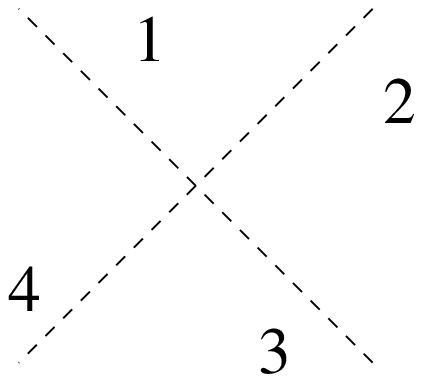}}
&\=&-\frac{\im}{2^3\cdot 4!}\,(k_1^2+3k_1\cdot k_3)\,F(k_1,k_2,k_3,k_4)\quad,\\
\parbox{2cm}{\includegraphics[width=1.9cm]{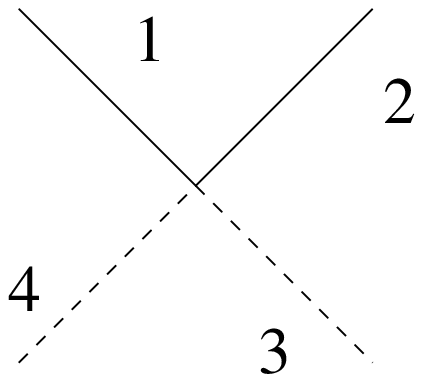}}
&\=&-\frac{\im}{2^5}\,(k_1^2-k_2^2+2 k_1\cdot k_3-2 k_2\cdot k_3 \non\\
&&\qquad\ +\,2k_1\wedge k_2+2k_1\wedge k_3+2k_3\wedge k_2)\,F(k_1,k_2,k_3,k_4)
\quad,
\ena
\end{itemize}
where we used the conventions of section 2 with the definitions
\beq
u\cdot v \= - \eta^{ab}\,u_a v_b   
\= u_tv_t - u_yv_y \qquad\text{and}\qquad
u\wedge v\=   u_tv_y-u_yv_t \quad.
\eeq
Moreover, we have defined
\beq
F(k_1,\dots, k_n) \= \exp \bigl\{ -\sfrac{\im}{2} 
\textstyle{\sum_{i<j}^n} k_i \wedge k_j \bigr\} \quad.
\eeq
and use the convention that all momentum lines are entering the
vertex and energy-momentum conservation has been taken into account. 

We now compute the scattering amplitudes 
$\varphi\varphi \to \varphi\varphi$, $\rho\rho \to \rho\rho$ and 
$\varphi\rho\to \varphi\rho$ 
and the production amplitude $\varphi\varphi \to \rho\rho$. 
We perform the calculations in the center-of-mass frame. We assign the
convention that particles 
with momenta $k_1$ and $k_2$ are incoming, while those with momenta 
$k_3$ and $k_4$ are outgoing.
  
\bigskip

\noindent
{\bf Amplitude $\varphi\varphi\to \varphi\varphi$.\ }
The four momenta are explicitly written as
\bea\label{momenta aa-aa}
k_1=(E,p)\ ,\quad k_2=(E,-p)\ ,\quad k_3=(-E,p)\ ,\quad k_4=(-E,-p)\ ,
\ena 
with the on-shell condition $E^2-p^2=4\a^2$.
There are two topologies of diagrams contributing to this process. 
Taking into account the leg permutations corresponding to the same 
particle at a single vertex, the contributions read\\
\begin{center}\begin{tabular}{r@{}clr@{}cl}
\parbox{1.9cm}{\includegraphics[width=1.9cm]{vertice1111.eps}}
&=&$2\im\a^2 \cos^2 (\theta Ep)\quad,$ & 
\parbox{2.4cm}{\includegraphics[width=2.4cm]{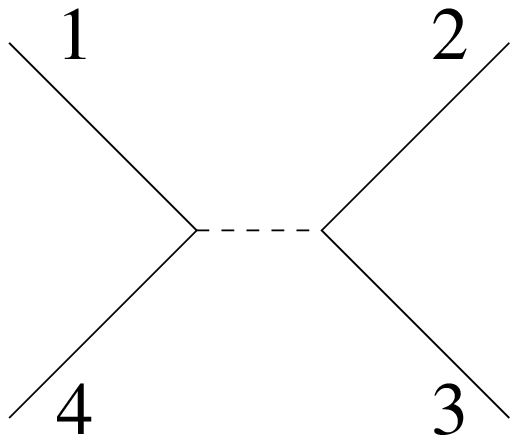}}
&=&$0\quad,$\\
\parbox{1.9cm}{\includegraphics[width=1.9cm]{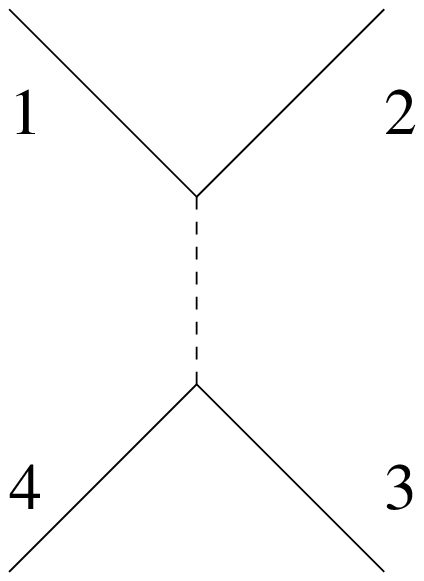}}
&=&$-\frac{\im}{2}p^2\sin^2 (\theta Ep)\quad,$ &
\parbox{2.4cm}{\includegraphics[width=2.4cm]{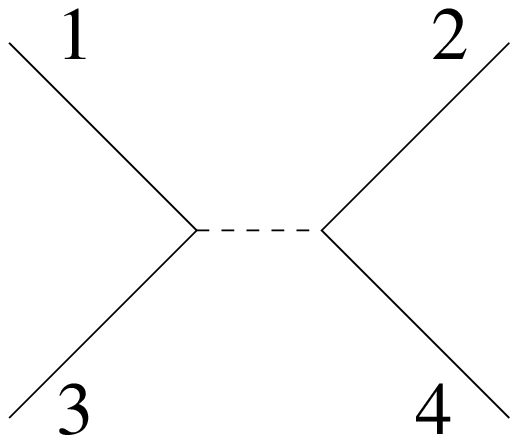}}
&=&$ \frac{\im}{2}E^2\sin^2 (\theta Ep)\quad.$ 
\end{tabular}\\ \end{center}
\noindent
The second diagram is actually affected by a collinear divergence
since the total momentum $k_1+k_4$ for the internal massless particle
is on-shell vanishing. 
We regularize this divergence by temporarily giving a small mass to
the $\rho$ particle. It is easy to see that the amplitude is 
zero for any value of the small mass since the wedge products $k_1 \wedge 
k_4$ and $k_2 \wedge k_3$ from the two vertices always vanish. 
As an alternative procedure we can put one of the external particles
slightly off-shell, so obtaining a finite result which vanishes in the 
on-shell limit. 

Summing all the contributions, for the $\varphi\varphi \to \varphi\varphi$ 
amplitude we arrive at 
\beq
A_{\varphi\varphi \to \varphi\varphi}\=2\im\a^2 \quad,
\eeq
which perfectly describes a {\em causal\/} amplitude.

A nonvanishing $\varphi\varphi \to \varphi\varphi$ amplitude appears also in 
the noncommutative sine-Gordon proposal of \cite{GP,GMPT}. However, there 
the amplitude has a nontrivial $\theta$-dependence which is responsible for 
acausal behavior. 
Comparing the present result with the result in \cite{GMPT}, we observe 
that the same kind of diagrams contribute. The main difference is that 
the exchanged particle is now massless instead of massive. 
This crucial difference leads to the cancellation of the 
$\theta$-dependent trigonometric behaviour which in the previous case
gave rise to acausality.

\bigskip

\noindent
{\bf Amplitude $\rho\rho \to \rho\rho$.\ }
In this case the center-of-mass momenta are given by
\beq
k_1=(E,E)\ ,\quad k_2=(E,-E)\ ,\quad k_3=(-E,E)\ ,\quad k_4=(-E,-E)\ ,
\label{rhomomenta}
\eeq  
where the on-shell condition $E^2-p^2=0$ has already been taken into account. 
For this amplitude we have the following contributions\\
\begin{center}\begin{tabular}{r@{}clr@{}cl}
\parbox{2cm}{\includegraphics[width=1.9cm]{vertice2222.eps}}
&=& $0\quad,$ &
\parbox{2.5cm}{\includegraphics[width=2.4cm]{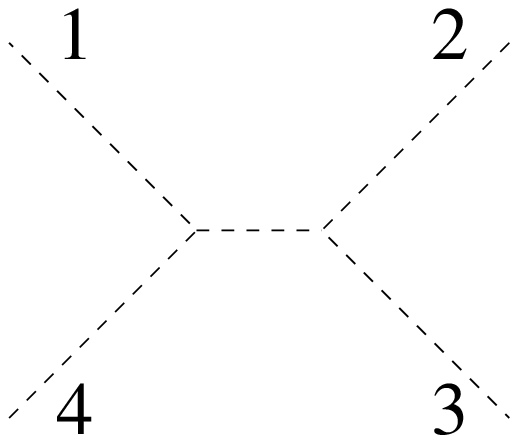}}
&=& $0\quad,$ \\
\parbox{2cm}{\includegraphics[width=1.9cm]{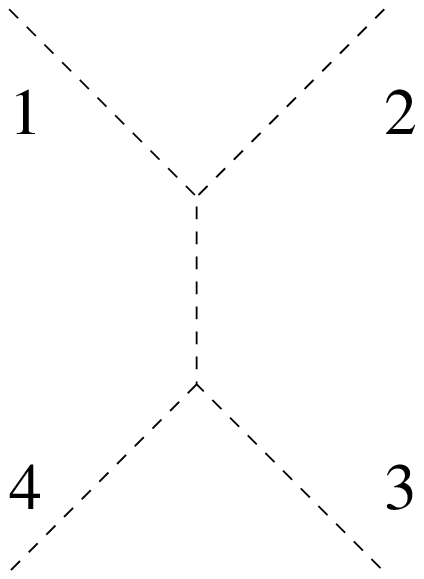}}
&=& $-\frac{\im}{2}E^2\sin^2 (\theta E^2)\quad,$ &
\parbox{2.5cm}{\includegraphics[width=2.4cm]{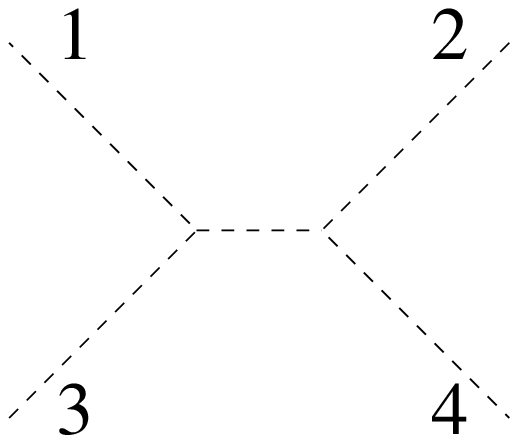}}
&=& $\frac{\im}{2}E^2\sin^2 (\theta E^2)\quad.$
\end{tabular}\\ \end{center}
Again, a collinear divergence appears in the second diagram.
In order to regularize the divergence we can proceed as before by
assigning a small mass to the $\rho$ particle. The main difference with respect
to the previous case is that now the $\rho$ particle also appears as an 
external particle, with the consequence that the on-shell momenta in 
(\ref{rhomomenta}) will get modified by the introduction of a regulator mass.
A careful calculation shows that the amplitude is zero for any value
of the regulator mass, due to the vanishing of the factors 
$k_1 \wedge k_4$ and $k_2 \wedge k_3$ from the vertices.

Therefore, the two nonvanishing contributions add to
\beq
A_{\rho\rho \to \rho\rho}\=0 \quad.
\eeq

\bigskip

\noindent
{\bf Amplitude $\varphi\rho \to \varphi\rho$.\ }
There are two possible 
configurations of momenta in the center-of-mass frame, describing the
scattering of the massive particle with either a left-moving or a right-moving
massless one. In the left-moving case the momenta are
\bea\label{momenta ab-ab 1} 
k_1=(E,p)\ ,\quad  k_2=(p,-p)\ ,\quad  k_3=(-E,p)\ , \quad  k_4=(-p,-p)\ ,
\ena
while in the right-moving case we have
\bea\label{momenta ab-ab 2} 
k_1=(E,-p)\ ,\quad k_2=(p,p)\ ,\quad k_3=(-E,p)\ , \quad  k_4=(-p,-p)\ .
\ena
For the left-moving case (\ref{momenta ab-ab 1}) the results are\\
\begin{center}
\begin{tabular}{r@{}cl@{\hspace{1cm}}r@{}cl}
\parbox{2cm}{\includegraphics[width=1.9cm]{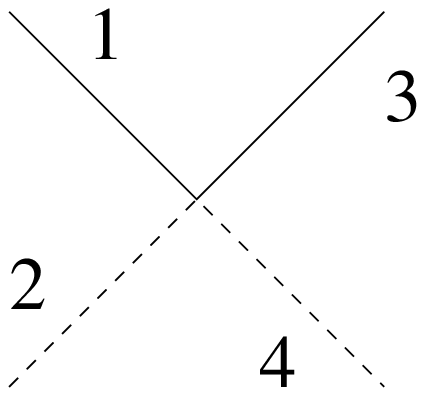}}
&=&\multicolumn{4}{l}{
$-\frac{\im}{2}Ep\,\sin(\theta Ep)\,\sin(\theta p^2)\quad,$}\\
\parbox{2cm}{\includegraphics[width=1.9cm]{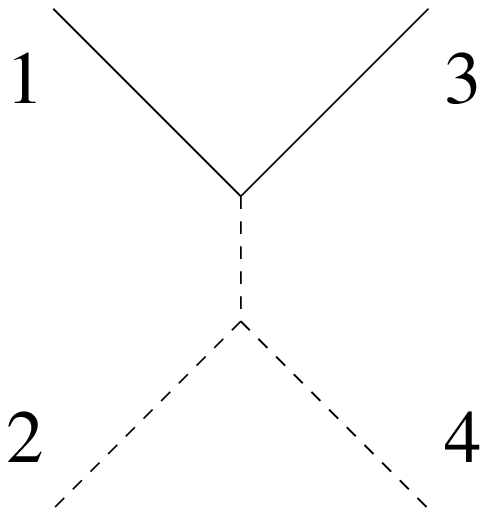}}
&=&\multicolumn{4}{l}{
$\frac{\im}{2}Ep\,\sin(\theta Ep)\,\sin(\theta p^2)\quad,$}\\
\parbox{2.5cm}{\includegraphics[width=2.4cm]{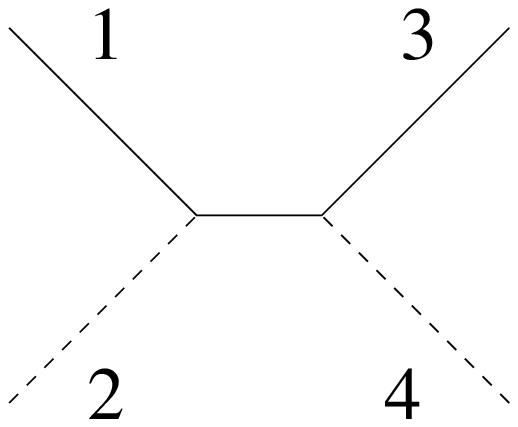}}
&=&$0\quad,$ &
\parbox{2.5cm}{\includegraphics[width=2.4cm]{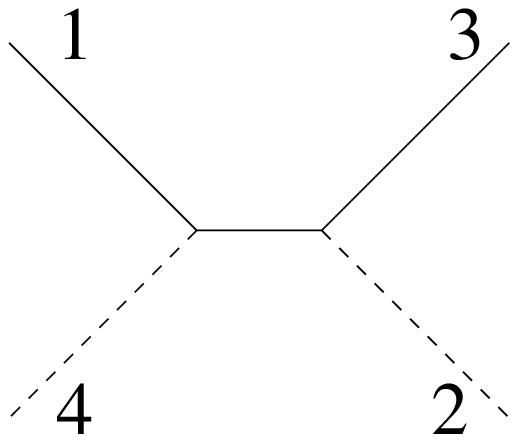}}
&=&$0\quad.$
\end{tabular}\\
\end{center}
For the right-moving choice (\ref{momenta ab-ab 2}), we obtain instead\\
\begin{center}
\begin{tabular}{r@{}cl@{\hspace{1cm}}r@{}cl}
\parbox{2cm}{\includegraphics[width=1.9cm]{varabab.eps}}
&=&$0\quad,$ &
\parbox{2cm}{\includegraphics[width=1.9cm]{varaapbb.eps}}
&=&$0\quad,$ \\
\parbox{2.5cm}{\includegraphics[width=2.4cm]{varabpab.eps}}
&=&$0\quad,$ &
\parbox{2.5cm}{\includegraphics[width=2.4cm]{abpab3.eps}}
&=&$0\quad.$
\end{tabular}\\ 
\end{center}
In this second case an infrared divergence is present due to the massless 
propagator, but again it can be cured as described before.
In both cases the scattering amplitude vanishes,
\beq
A_{\varphi\rho \to \varphi\rho}\=0 \quad.
\eeq

\bigskip

\noindent
{\bf Amplitude $\varphi\varphi \to \rho\rho$.\ }
The momenta in the center-of-mass frame are given by
\bea\label{momenta aa-bb}
k_1=(E,p)\ ,\quad k_2=(E,-p)\ ,\quad k_3=(-E,E)\ ,\quad k_4=(-E,-E)\ .
\ena 
In this case we have three kinds of diagrams contributing. 
The corresponding results are\\
\begin{center}
\begin{tabular}{r@{}cl@{\hspace{1cm}}r@{}cl}
\parbox{2cm}{\includegraphics[width=1.9cm]{vertice1122.eps}}
&=& \multicolumn{4}{l}{
$\frac{\im}{2}Ep\,\sin(\theta Ep)\,\sin(\theta E^2)\quad,$} \\
\parbox{2cm}{\includegraphics[width=1.9cm]{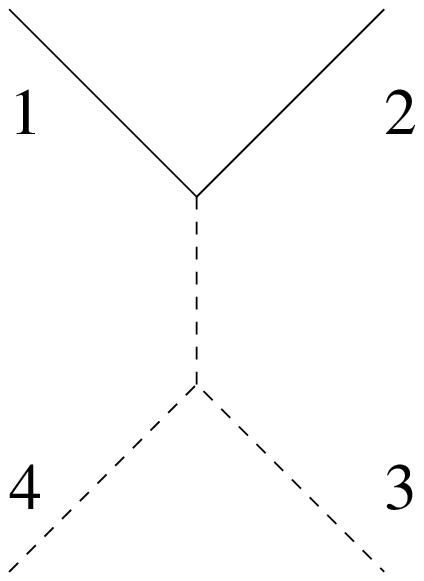}}
&=& \multicolumn{4}{l}{
$-\frac{\im}{2}Ep\,\sin(\theta Ep)\,\sin(\theta E^2)\quad,$} \\
\parbox{2.5cm}{\includegraphics[width=2.4cm]{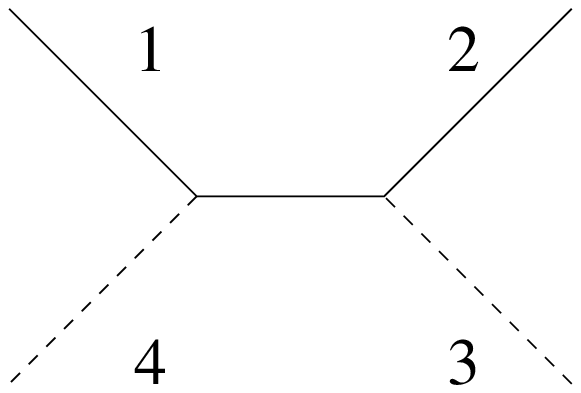}}
&=&$0\quad,$ &
\parbox{2.5cm}{\includegraphics[width=2.4cm]{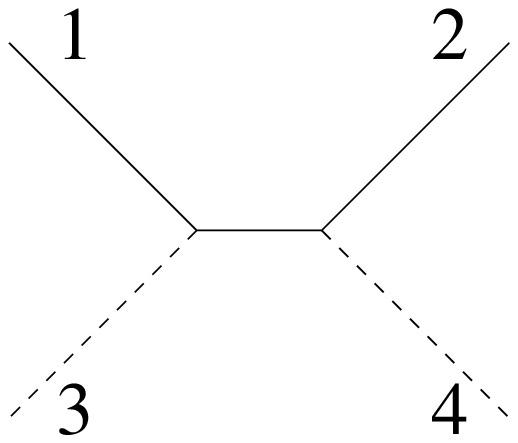}}
&=&$0\quad.$
\end{tabular}\\
\end{center}
\noindent
Summing the four contributions, we obtain
\beq
A_{\varphi\varphi \to \rho\rho}\=0
\eeq
as it should be expected for a production amplitude in an integrable model.
The same is true for the time-reversed production,
\beq
A_{\rho\rho \to \varphi\varphi}\=0 \quad.
\eeq

Summarizing, we have found 
that the only nonzero amplitude for tree-level $2 \to 2$ processes 
is the one describing the scattering among two of the massive excitations. 
The result is constant, independent of the momenta and so describes a 
perfectly {\em causal\/} process.
Since the result is independent of the noncommutation parameter $\theta$ 
it agrees with the four-point amplitude for the ordinary sine-Gordon model.
Finally, we have found that the production amplitudes 
$\varphi \varphi \to \rho \rho$ and $\rho\rho \to \varphi\varphi$ vanish, 
as required for ordinary integrable theories.

As a further check of our calculation and an additional test of our model  
we have computed the production amplitude 
$\varphi \varphi \to \varphi \varphi \varphi \varphi$ 
and the scattering amplitude 
$\varphi \varphi \varphi \to \varphi \varphi \varphi$. 
In both cases the topologies we have to consider are
\vspace{0.5cm}
\begin{center}
\begin{tabular}{c@{\hspace{0.8cm}}c@{\hspace{0.8cm}}c}
\parbox{2.5cm}{\includegraphics[width=2.4cm]{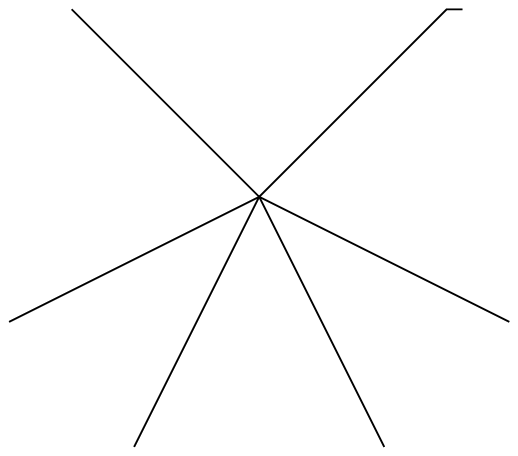}}&
\parbox{2.3cm}{\includegraphics[width=2.2cm]{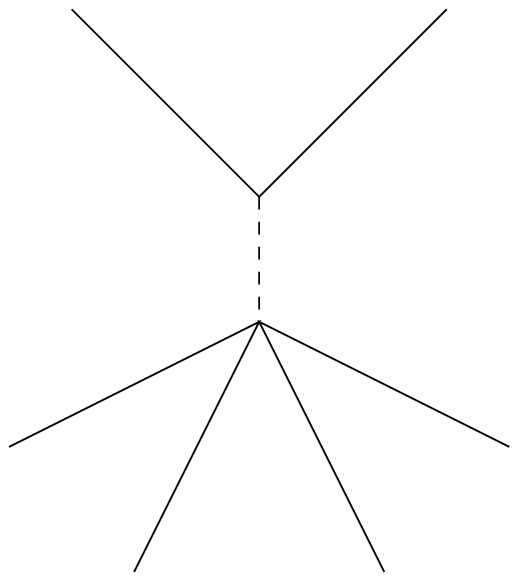}}& 
\parbox{2.0cm}{\includegraphics[width=1.9cm]{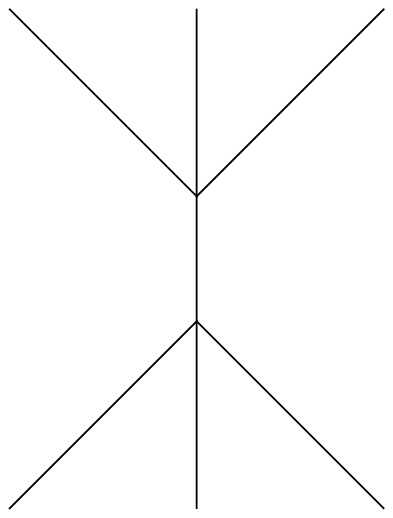}} \end{tabular} \\
\begin{tabular}{c@{\hspace{0.8cm}}c@{\hspace{0.8cm}}c@{\hspace{0.8cm}}c}
\parbox{2.0cm}{\includegraphics[width=1.9cm]{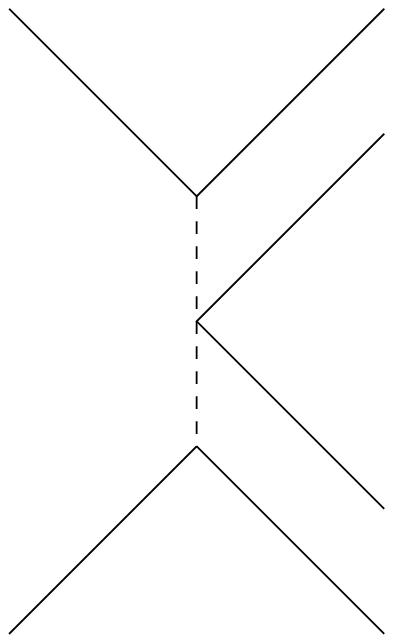}}&
\parbox{2.0cm}{\includegraphics[width=1.9cm]{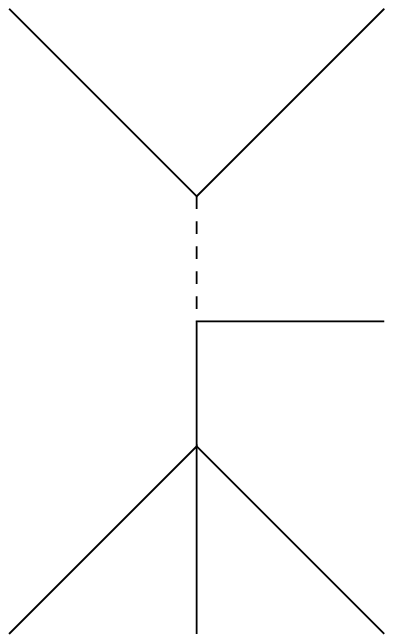}}&
\parbox{2.0cm}{\includegraphics[width=1.9cm]{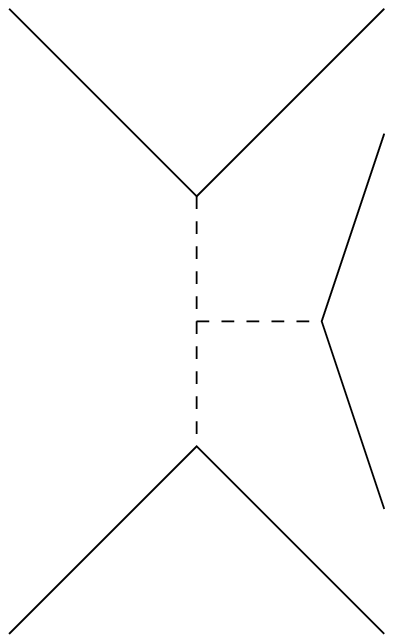}}&
\parbox{1.8cm}{\includegraphics[width=1.7cm]{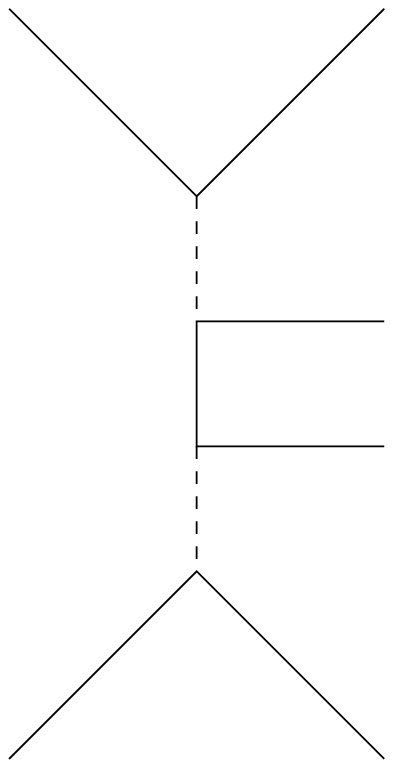}}\quad.
\end{tabular}\\ \end{center}
\vspace{0.5cm}
\noindent
Due to the growing number of channels and ordering of vertices, it is
no longer practical to perform the calculations by hand.
We have used {\it Mathematica}$^{\copyright}$ 
to symmetrize the vertices and take automatically into account the 
different diagrams obtained by exchanging momenta entering a given vertex.
The computation has been performed with assigned values of the external
momenta but arbitrary values for $\alpha^2$ and $\theta$. 
We have found a vanishing result for both the 
scattering and the production amplitude. 
This is in agreement with the commutative sine-Gordon model results.

\bigskip

\noindent
{\bf Amplitudes in the ``$h$-model''.\ }
We now discuss the $2\to 2$ amplitudes in the Leznov formulation. 
The theory is again described by two interacting fields, $h_1$ 
massless and $h_2$ massive. Referring to the action (\ref{haction})
we extract the following Feynman rules,
\begin{itemize}
\item{The propagators
\bea
\parbox{2cm}{\includegraphics[width=1.9cm]{prop2.eps}}
&\ \equiv\ &\langle h_1 h_1 \rangle\=\frac{\im}{2k^2}\quad,\\
\parbox{2cm}{\includegraphics[width=1.9cm]{prop1.eps}}
&\ \equiv\ &\langle h_2 h_2 \rangle\=\frac{\im/2}{k^2-4\a^2}\quad.
\ena}
\item{The vertex
\bea
\parbox{2cm}{\includegraphics[width=1.9cm]{vertice112.eps}}
&\=&- 4\a\,(k_{3t}-k_{3y})\, F(k_1,k_2,k_3)\quad.
\ena}
\end{itemize}
Again, we compute scattering amplitudes in the center-of-mass frame.
Given the particular structure of the vertex, at tree level there
is no $h_1 h_1 \to h_1 h_1$ scattering.
To find the $h_2 h_2 \to h_2 h_2$ amplitude we assign the momenta
(\ref{momenta aa-aa}) to the external particles. 
The contributions are \\
\begin{center}\begin{tabular}{r@{}clr@{}cl}
\parbox{2.3cm}{\includegraphics[width=2.2cm]{aapaa2.eps}}
&=& $-16\im\a^2\cos^2(\theta Ep)\quad,$ &
\parbox{2cm}{\includegraphics[width=1.9cm]{11p11fig.eps}}
&=& $16\im\a^2\cos^2(\theta Ep)\quad,$ \\
\parbox{2.3cm}{\includegraphics[width=2.2cm]{aapaa3.eps}}
&=&$0\quad.$ 
\end{tabular}\\ 
\end{center}
\noindent
We note that a collinear divergence appears in the last diagram which
can be regularized as described before.
Summing the two nonvanishing contributions we obtain complete cancellation.

For the $h_2 h_2 \to h_1 h_1$ amplitude
the center-of-mass-momenta are given in (\ref{momenta aa-bb}).
The only topology contributing to this production amplitude has 
two channels, yielding \\
\begin{center}\begin{tabular}{r@{}clr@{}cl}
\parbox{2.5cm}{\includegraphics[width=2.4cm]{12p12fig.eps}}
&=&$0\quad,$ &
\parbox{2.3cm}{\includegraphics[width=2.2cm]{abpab2.eps}}
&=& $0\quad,$
\end{tabular}\\ 
\end{center}
\noindent 
which are both zero, so giving a vanishing result once more.
The same is true for the $h_1 h_1 \to h_2 h_2$ production process.

Finally, for the $h_1 h_2 \to h_1 h_2$ amplitude, we refer to the 
center-of-mass momenta defined in (\ref{momenta ab-ab 1}) 
and (\ref{momenta ab-ab 2}). In both cases the contributions are\\
\begin{center}\begin{tabular}{r@{}clr@{}cl}
\parbox{2.3cm}{\includegraphics[width=2.2cm]{varabpab.eps}}
&=&$0\quad,$ &
\parbox{2.3cm}{\includegraphics[width=2.2cm]{abpab3.eps}}
&=&$0\quad,$
\end{tabular}\\ \end{center}
\noindent 
and so we find that the sum of the two channels is always equal to zero.

Since all the $2 \to 2$ amplitudes vanish, the S-matrix is trivially causal
and factorized.

Both in the ordinary and noncommutative cases the ``$h$-model''  is dual to 
the ``$g$-model''. In the commutative limit the ``$g$-model'' gives rise to a 
sine-Gordon model plus a free field which can be set to zero. 
In this limit our amplitudes exactly reproduce the  sine-Gordon amplitudes. 
On the other hand, the amplitudes for the ``$h$-model'' all vanish. Therefore,
in the commutative limit they do not reproduce anything immediately 
recognizable as an ordinary sine-Gordon amplitude. This can be understood by 
observing that, both in the ordinary and in the noncommutative case, the
Leznov formulation is an alternative description of the sine-Gordon dynamics
and obtained from the standard Yang formulation by the {\em nonlocal field 
redefinition\/} given in~(\ref{nonlocal3}). Therefore, it is expected
that the scattering amplitudes for the elementary exitations, which are
different in the two formulations, do not resemble each other.

\section{Conclusions}

We have proposed a novel noncommutative sine-Gordon system based on
{\em two\/} scalar fields, which seems to retain all advantages of
$1{+}1$ dimensional integrable models known from the commutative limit.
The rationale for introducing a second scalar field was provided by
deriving the sine-Gordon equations and action through dimensional and 
algebraic reduction of an integrable $2{+}1$ dimensional sigma model: 
In the noncommutative extension of this scheme it is natural to generalize 
the algebraic reduction of SU(2)$\to$U(1) to one of U(2)$\to$U(1)$\times$U(1).
We gave two Yang-type and one Leznov-type parametrizations of the coupled 
system in (\ref{Yphi}), (\ref{Yrho}) and (\ref{Lh12}) and provided the actions
for them, including a comparison with previous proposals.
It was then outlined how to explicitly construct noncommutative sine-Gordon
multi-solitons via the dressing method based on the underlying linear system.
We found that the one-soliton configuration agrees with the commutative one
but already the two-soliton solutions gets Moyal deformed.

What is the gain of doubling the field content as compared to the standard
sine-Gordon system or its straightforward star deformation?
Usually, time-space noncommutativity adversely affects the causality and
unitarity of the S-matrix (see, e.g.~\cite{CM, GP, GMPT}), even in the
presence of an infinite number of local conservation laws. In contrast,
the model described here seems to possess an S-matrix which is {\em causal\/} 
and {\em factorized\/}, as we checked for all tree-level $2\to 2$ processes
both in the Yang and Leznov formulations. Furthermore, we verified the
vanishing of some $3\to 3$ scattering amplitudes and $2\to 4$ production
amplitudes thus proving the absence of particle production.

It would be nice to understand what actually drives a system to be 
integrable in the noncommutative case. A hint in this direction might
be that the model proposed in~\cite{GP} has been constructed directly in 
two dimensions even if its equations of motion (but {\em not\/} the
action) can be obtained by a suitable reduction of a four dimensional
system (noncommutative self-dual Yang-Mills). The model proposed in this 
paper, instead, originates directly, already at the level of the action,
from the reduction of noncommutative self-dual Yang-Mills theory 
which is known to be integrable and related to the $N{=}2$ string~\cite{LPS}. 

Several directions of future research are suggested by our results.
First, one might hope that our noncommutative two-field sine-Gordon model
is equivalent to some two-fermion model via noncommutative bosonization.
Second, it would be illuminating to derive the exact two-soliton solution
and extract its scattering properties, either directly in our model or by
reducing wave-like solutions of the 2+1 dimensional sigma model 
\cite{bieling,wolf}.
Third, there is no obstruction against applying the ideas and techniques 
of this paper to other 1+1 dimensional noncommutative integrable systems 
in order to cure their pathologies as well.

\bigskip

\bigskip

\noindent
{\bf Acknowledgements}

L.M. and L.T. acknowledge a useful discussion with G. Mussardo.
This work was partially supported by the Deutsche Forschungsgemeinschaft 
(DFG), INFN, MURST and the European Commission RTN program
HPRN--CT--2000--00131, in which S.P. and L.M. are associated to
the University of Padova. 

\bigskip

\bigskip


\begin{thebibliography}{99}

\bibitem{SW}
N.~Seiberg and E.~Witten,
{\sl JHEP} {\bf 9909} (1999) 032 [hep-th/9908142].

\bibitem{LPS} 
O.~Lechtenfeld, A.D.~Popov and B.~Spendig,
{\sl Phys.\ Lett.} {\bf B 507} (2001) 317 [hep-th/0012200].

\bibitem{T} 
K.~Takasaki, 
{\sl J.\ Geom.\ Phys.} {\bf 37} (2001) 291 [hep-th/0005194].

\bibitem{dimred} 
R.S. Ward,
{\sl Phil.\ Trans.\ R.\ Soc.\ Lond.} {\bf A 315} (1985) 451;\\
L.J. Mason and G.A.J. Sparling,
{\sl Phys.\ Lett.} {\bf A 137} (1989) 29; 
{\sl J.\ Geom.\ Phys.} {\bf 8} (1992) 243; \\
M.J. Ablowitz, S. Chakravarty and L.A. Takhtajan, 
{\sl Math.\ Phys.} {\bf 158} (1993), 289; \\
T.A. Ivanova and A.D. Popov, 
{\sl Phys.\ Lett.} {\bf A 205} (1995) 158 [hep-th/9508129];\\ 
{\sl Theor. Math. Phys.} {\bf 102} (1995) 280
[{\sl Teor. Mat. Fiz.} {\bf 102} (1995) 384]; \\  
S. Chakravarty, S.L. Kent and E.T. Newman, 
{\sl J. Math. Phys.} {\bf 36} (1995) 763; \\ 
M. Legare, {\sl Int.\ J.\ Mod.\ Phys.} {\bf A 12} (1997) 219.

\bibitem{reviews}
M.R.~Douglas and N.A.~Nekrasov,
{\sl Rev. Mod. Phys.} {\bf 73} (2002) 977 [hep-th/0106048];\\
A.~Konechny and A.~Schwarz,
{\sl Phys.\ Rept.} {\bf 360} (2002) 353       [hep-th/0107251];\\
R.J.~Szabo,
{\sl Phys.\ Rept.} {\bf 378} (2003) 207 [hep-th/0109162];\\
M.~Hamanaka,
{\em ``Noncommutative solitons and D-branes,''} hep-th/0303256.

\bibitem{Legare} 
M.~Legar\'e, 
{\em ``Noncommutative generalized NS and super matrix KdV systems from a 
noncommutative version of (anti-)self-dual Yang-Mills equations,''} 
hep-th/0012077.

\bibitem{LPS2} 
O.~Lechtenfeld, A.D.~Popov and B.~Spendig,
{\sl JHEP} {\bf 0106} (2001) 011 [hep-th/0103196].

\bibitem{LP1} 
O.~Lechtenfeld and A.D.~Popov,
{\sl JHEP} {\bf 0111} (2001) 040 [hep-th/0106213];\\
{\sl Phys.\ Lett.} {\bf B 523} (2001) 178 [hep-th/0108118].

\bibitem{bieling}
S.~Bieling,
{\sl J. Phys.} {\bf A 35} (2002) 6281 [hep-th/0203269].

\bibitem{wolf}
M.~Wolf,
{\sl JHEP} {\bf 0206} (2002) 055 [hep-th/0204185];\\
M.~Ihl and S.~Uhlmann,
{\sl Int.\ J.\ Mod.\ Phys.} {\bf A 18} (2003) 4889 [hep-th/0211263].

\bibitem{goteborg}
O.~Lechtenfeld,
{\sl Fortschr. Phys.} {\bf 52} (2004) 596 [hep-th/0401158].

\bibitem{list1}
D.~Bak,
{\sl Phys.\ Lett.} {\bf B 471} (1999) 149 [hep-th/9910135]; \\
D.J.~Gross and N.A.~Nekrasov,
{\sl JHEP} {\bf 0007} (2000) 034 [hep-th/0005204]; \\
{\sl JHEP} {\bf 0103} (2001) 044 [hep-th/0010090]; \\
A.P.~Polychronakos,
{\sl Phys.\ Lett.} {\bf B 495} (2000) 407 [hep-th/0007043]; \\
M.~Hamanaka and S.~Terashima,
{\sl JHEP} {\bf 0103} (2001) 034 [hep-th/0010221]; \\
K.~Hashimoto,
{\sl JHEP} {\bf 0012} (2000) 023 [hep-th/0010251]; \\
O.~Lechtenfeld and A.D.~Popov,
{\sl JHEP} {\bf 0401} (2004) 069 [hep-th/0306263]. 

\bibitem{DMH}
A.~Dimakis and F.~M\"uller-Hoissen,\\
{\em ``A noncommutative version of the nonlinear Schr\"odinger equation,''}
hep-th/0007015;\\
{\sl J.\ Phys.} {\bf A 34} (2001) 9163 [nlin.si/0104071];
{\sl J.\ Phys.} {\bf A 37} (2004) 4069 [hep-th/0401142].

\bibitem{list2}
L.D.~Paniak,
{\em ``Exact Noncommutative KP and KdV Multi-solitons,''} hep-th/0105185;\\
N.~Wang and M.~Wadati,
{\sl J.\ Phys.\ Soc.\ Jap.} {\bf 72} (2003) 1366; 1881; 3055.

\bibitem{murugan}
B.H.~Lee, K.M.~Lee and H.S.~Yang,
{\sl Phys.\ Lett.} {\bf B 498} (2001) 277 [hep-th/0007140];\\
K.~Furuta, T.~Inami, H.~Nakajima and M.~Yamamoto,\\
{\sl Phys.\ Lett.} {\bf B 537} (2002) 165 [hep-th/0203125];\\
J.~Murugan and R.~Adams,
{\sl JHEP} {\bf 0212} (2002) 073 [hep-th/0211171];\\
H.~Otsu, T.~Sato, H.~Ikemori and S.~Kitakado,
{\sl JHEP} {\bf 0307} (2003) 054 [hep-th/0303090];\\
A.~Hanany and D.~Tong,
{\sl JHEP} {\bf 0307} (2003) 037 [hep-th/0306150];\\
J.~Murugan and A.~Millner,
{\em ``Transmogrifying fuzzy vortices,''}
hep-th/0403105.

\bibitem{hamanaka}
M.~Hamanaka and K.~Toda,
{\sl J.\ Phys.} {\bf A 36} (2003) 11981 [hep-th/0301213];\\
{\em ``Towards noncommutative integrable equations,''}
hep-th/0309265;\\
M.~Hamanaka,
{\em ``Commuting flows and conservation laws for noncommutative
       Lax hierarchies,''}
hep-th/0311206;\\
K.M.~Lee,
{\em ``Chern-Simons solitons, chiral model, and (affine) Toda model on
       noncommutative space,''}
hep-th/0405244.

\bibitem{GMPT} 
M.T. Grisaru, L. Mazzanti, S. Penati and L. Tamassia,\\
{\sl JHEP} {\bf 0404} (2004) 057 [hep-th/0310214].

\bibitem{seiberg} 
N. Seiberg, L. Susskind and N. Toumbas, 
{\sl JHEP} {\bf 0006} (2000) 044 [hep-th/0005015].

\bibitem{gm} 
J. Gomis and T. Mehen, 
{\sl Nucl.\ Phys.} {\bf B 591} (2000) 265 [hep-th/0005129].

\bibitem{AG}
L.~Alvarez-Gaume, J.L.F.~Barbon and R.~Zwicky,
{\sl JHEP} {\bf 0105} (2001) 057 [hep-th/0103069].

\bibitem{CM} 
I. Cabrera-Carnero and M. Moriconi,
{\sl Nucl.\ Phys.} {\bf B 673} (2003) 437 [hep-th/0211193].

\bibitem{GP} 
M.T. Grisaru and S. Penati,
{\sl Nucl.\ Phys.} {\bf B 655} (2003) 250 [hep-th/0112246].

\bibitem{ward}
R.S.~Ward,
{\sl J. Math. Phys.} {\bf 29} (1988) 386;
{\sl Commun. Math. Phys.} {\bf 128} (1990) 319.

\bibitem{Y}
C.N.~Yang,
{\sl Phys. Rev. Lett.} {\bf 38} (1977) 1377.

\bibitem{L}
A.N.~Leznov,
{\sl Theor. Math. Phys.} {\bf 73} (1987) 1233.

\bibitem{dressing}
V.E.~Zakharov and A.V.~Mikhailov,
{\sl Sov. Phys. JETP} {\bf 47} (1978) 1017;\\
V.E.~Zakharov and A.B.~Shabat,
{\sl Funct. Anal. Appl.} {\bf 13} (1979) 166;\\
P.~Forg\'acs, Z.~Horv\'ath and L.~Palla,
{\sl Nucl. Phys.} {\bf B 229} (1983) 77.

\bibitem{faddeev}
L.D.~Faddeev and L.A.~Takhtajan,
{\em ``Hamiltonian methods in the theory of solitons,''}\\
Springer 1987.

\bibitem{ivle1}
T.A.~Ivanova and O.~Lechtenfeld,
{\sl Int. J. Mod. Phys.} {\bf A 16} (2001) 303 [hep-th/0007049].

\bibitem{ioannidou}
T.A.~Ioannidou,
{\sl J.\ Math.\ Phys.} {\bf 37} (1996) 3422 [hep-th/9604126]; \\
T.A.~Ioannidou and W.J.~Zakrzewski,
{\sl J.\ Math.\ Phys.} {\bf 39} (1998) 2693 [hep-th/9802122].

\bibitem{nair}
V.P.~Nair and J.~Schiff,
{\sl Nucl.\ Phys.} {\bf B 371} (1992) 329.

\bibitem{moore}
A.~Losev, G.W.~Moore, N.~Nekrasov and S.~Shatashvili,\\
{\sl Nucl.\ Phys.\ Proc.\ Suppl.} {\bf 46} (1996) 130 [hep-th/9509151].

\bibitem{witten}
E. Witten,
{\sl Commun.\ Math.\ Phys.} {\bf 92} (1984) 455.

\bibitem{bosonization}
A.M. Polyakov and P.B. Wiegmann,\\
{\sl Phys.\ Lett.} {\bf B 131} (1983) 121;
{\sl Phys.\ Lett.} {\bf B 141} (1984) 223;\\
P. Di Vecchia and P. Rossi,
{\sl Phys.\ Lett.} {\bf B 140} (1984) 344;\\{}
P. Di Vecchia, B. Durhuus and J.L. Petersen,
{\sl Phys.\ Lett.} {\bf B 144} (1984) 245;\\{}
Y.~Frishmann, 
{\sl Phys.\ Lett.} {\bf B 146} (1984) 204; \\
P.~Goddard and D.~Olive, 
{\sl Int.\ Journ.\ Mod.\ Phys.} {\bf A 1} (1986) 303;\\
C.P. Burgess and F. Quevedo,
{\sl Phys.\ Lett.} {\bf B 329} (1994) 457 [hep-th/9403173].

\bibitem{bosonization2}
J.C. Le Guillou, E. Moreno, C. Nunez and F.A. Schaposnik,\\
{\sl Nucl.\ Phys.} {\bf B 484} (1997) 682 [hep-th/9609202].

\bibitem{MS}
E.F. Moreno and F.A. Schaposnik, 
{\sl Nucl.\ Phys.} {\bf B 596} (2001) 439 [hep-th/0008118];\\
{\sl JHEP} {\bf 0003} (2000) 032 [hep-th/0002236].

\bibitem{matsubara}
K. Matsubara,
{\sl Phys.\ Lett.} {\bf B 482} (2000) 417 [hep-th/0003294].

\bibitem{mass}
C.P. Burgess and F. Quevedo,
{\sl Nucl.\ Phys.} {\bf B 421} (1994) 373 [hep-th/9401105];\\
C. Nunez, K. Olsen and R. Schiappa,
{\sl JHEP} {\bf 0007} (2000) 030 [hep-th/0005059].

\end{thebibliography}
\end{document}